\documentclass[prl,twocolumn,letterpaper, superscriptaddress]{revtex4-1}
\usepackage{graphicx}
\usepackage{dcolumn}
\usepackage{bm}
\usepackage{color}
\usepackage{tabularx}
\usepackage{array}
\usepackage{amsmath}
\usepackage{stmaryrd}  

\begin{document}

\title{Coherent coupling of two remote magnonic resonators mediated by superconducting circuits}

\author{Yi Li}
\affiliation{Materials Science Division, Argonne National Laboratory, Argonne, IL 60439, USA}

\author{Volodymyr G. Yefremenko}
\affiliation{High Energy Physics Division, Argonne National Laboratory, Lemont, IL 60439, USA}

\author{Marharyta Lisovenko}
\affiliation{High Energy Physics Division, Argonne National Laboratory, Lemont, IL 60439, USA}

\author{Cody Trevillian}
\affiliation{Department of Physics, Oakland University, Rochester, Michigan 48309, USA}

\author{Tomas Polakovic}
\affiliation{Physics Division, Argonne National Laboratory, Lemont, IL 60439, USA}

\author{Thomas W. Cecil}
\affiliation{High Energy Physics Division, Argonne National Laboratory, Lemont, IL 60439, USA}

\author{Pete S. Barry}
\affiliation{High Energy Physics Division, Argonne National Laboratory, Lemont, IL 60439, USA}

\author{John Pearson}
\affiliation{Materials Science Division, Argonne National Laboratory, Argonne, IL 60439, USA}

\author{Ralu Divan}
\affiliation{Center for Nanoscale Materials, Argonne National Laboratory, Argonne, IL 60439, USA}

\author{Vasyl Tyberkevych}
\affiliation{Department of Physics, Oakland University, Rochester, Michigan 48309, USA}

\author{Clarence L. Chang}
\affiliation{High Energy Physics Division, Argonne National Laboratory, Lemont, IL 60439, USA}

\author{Ulrich Welp}
\affiliation{Materials Science Division, Argonne National Laboratory, Argonne, IL 60439, USA}

\author{Wai-Kwong Kwok}
\affiliation{Materials Science Division, Argonne National Laboratory, Argonne, IL 60439, USA}

\author{Valentine Novosad}
\email{novosad@anl.gov}
\affiliation{Materials Science Division, Argonne National Laboratory, Argonne, IL 60439, USA}

\date{\today}

\begin{abstract}

We demonstrate microwave-mediated distant magnon-magnon coupling on a superconducting circuit platform, incorporating chip-mounted single-crystal Y$_3$Fe$_5$O$_{12}$ (YIG) spheres. Coherent level repulsion and dissipative level attraction between the magnon modes of the two YIG spheres are demonstrated. The former is mediated by cavity photons of a superconducting resonator, and the latter is mediated by propagating photons of a coplanar waveguide. Our results open new avenues towards exploring integrated hybrid magnonic networks for coherent information processing on a quantum-compatible superconducting platform.

\end{abstract}

\maketitle

The mainstream developments in quantum information processing rely on hybrid dynamic systems which combine different quantum modules for complementary functionality \cite{KurizkiPNAS2015,ClerkNPhys2020}. Among different hybrid systems, superconducting integrated circuits \cite{WallraffNature2004} are special because of the ease of scaling up to multi-qubit networks and being integrated to other quantum modules. In particular, the framework of superconducting circuits has been applied for bridging different physical degrees of freedom, such as optic photons \cite{ZhuNatureNano2018,FanSciAdv2018}, phonons \cite{ChuScience2017,SatzingerNature2018}, spins \cite{SchusterPRL2010,KuboPRL2010} and magnons \cite{LiPRL2019_magnon,HouPRL2019,McKenziePRB2019}.

One of the rapid growing subfields in hybrid systems is cavity magnonics \cite{LachanceQuirionAPEx2019,LiJAP2020,RameshtiarXiv2021}, which emphasizes on cavity-enhanced interactions between magnons and photons \cite{SoykalPRL2010,HueblPRL2013,TabuchiPRL2014,ZhangPRL2014,BaiPRL2015,KuzmichevJETP2020}. Magnons, solid-state excitations of spins, have been extensively explored for wave-based computing concepts \cite{ChumakNPhys2015}. Their collective dynamics allow for drastically enhanced magnetic dipolar interactions and thus strong coupling with photons, in comparison with diluted spins \cite{ImamogluPRL2009,WesenbergPRL2009}. The recent demonstration of magnon-qubit entanglement \cite{TabuchiScience2015,LachanceQuirionScience2020} has further triggered the pursuit of quantum operations with magnons \cite{TrifunovicPRX2013,AndrichnpjQuantumInf2017,RusconiPRA2019,ElyasiPRB2020,NeumanPRL2020}.

Increasing efforts have been devoted to chip-embedded cavity magnonic circuits with superconducting resonators \cite{HueblPRL2013,LiPRL2019_magnon,HouPRL2019,MorrisSREP2017,MandalAPL2020,HaygoodPRApplied2021,BaityAPL2021} in order to utilize their high quality factors and connectivity to qubits. However, the progression to multiple-device hybrid magnonic networks has been slow. One challenge is the degradation of magnon and photon coherences when they are integrated together. Examples include microwave quality factor reduction due to impedance mismatch from magnetic device mounting \cite{LiPRL2019_magnon,HouPRL2019}, increased damping for free-standing magnetic devices \cite{BaityAPL2021,HeyrothPRApplied2019,TremplerAPL2020}, and excitation of nonuniform magnon modes \cite{MorrisSREP2017}. Furthermore, low-damping YIG thin films, which are ideal for device integration, are typically grown on Gd$_3$Ga$_5$O$_{12}$ (GGG) substrates exhibiting excessive microwave losses at cryogenic temperature \cite{KosenAPLMaterials2019}. Thus it is highly desirable to develop a cryogenic circuit platform with a smooth integration of long-coherence magnonic systems in superconducting circuits.

In this work, we demonstrate remote magnon-magnon coupling in a compact superconducting-magnon hybrid circuit, using single-crystal YIG spheres that are mounted in lithographically defined holes on silicon substrates with superconducting resonators. The all-lithographic circuit design allows for arbitrary engineering of hybrid magnonic dynamics while achieving long magnon coherence. For a single 250-$\mu$m-diameter YIG sphere, we achieve a magnon-photon coupling strength of 130 MHz and a cooperativity of 13000 with both the magnon and photon damping rates approaching 1 MHz at 1.6 K. For a two-sphere-one-resonator circuit, we achieve a resonator mediated magnon-magnon coupling strength from 10 to 40 MHz in the dispersive coupling regime. Furthermore, we also achieve dissipative coupling \cite{HarderPRL2018,BhoiPRB2019,BoventerPRR2020,TserkovnyakPRR2020} of the magnon modes between the two YIG spheres mediated by propagating microwave photons. Both the coherent and dissipative magnon-magnon coupling strengths can be quantitatively reproduced by microwave circuit modeling. Our results provide a feasible platform for studying hybrid magnonic quantum networks at cryogenic temperatures \cite{RusconiPRA2019}.

Shown in Fig. 1(a), the design of the superconducting resonator (pink) resembles a $\lambda$/4 resonator with a circular loop antenna at the current antinode. A YIG sphere is mounted at the center of the loop which couples to the uniform (Kittel) magnon mode of the sphere. The superconducting circuits were fabricated from 200-nm NbN film sputtered on a high-resistance Si substrate. Photolithography and reactive ion etching were used for defining the circuits. Subsequently, circular patterns were lithographically defined on the Si substrate at the center of the loop antennas, where the superconducting film has been etched. Deep holes were etched on Si with a depth of 125 $\mu$m and a diameter of 250 $\mu$m, matching the dimension of the YIG spheres. For YIG sphere mounting, a tiny drop of diluted GE varnish was applied to each hole, and the YIG sphere was then quickly placed in the hole (Fig. 1b). The mounted circuits were placed in air for 24 hour for the GE varnish to dry out.

\begin{figure}[htb]
 \centering
 \includegraphics[width=3.0 in]{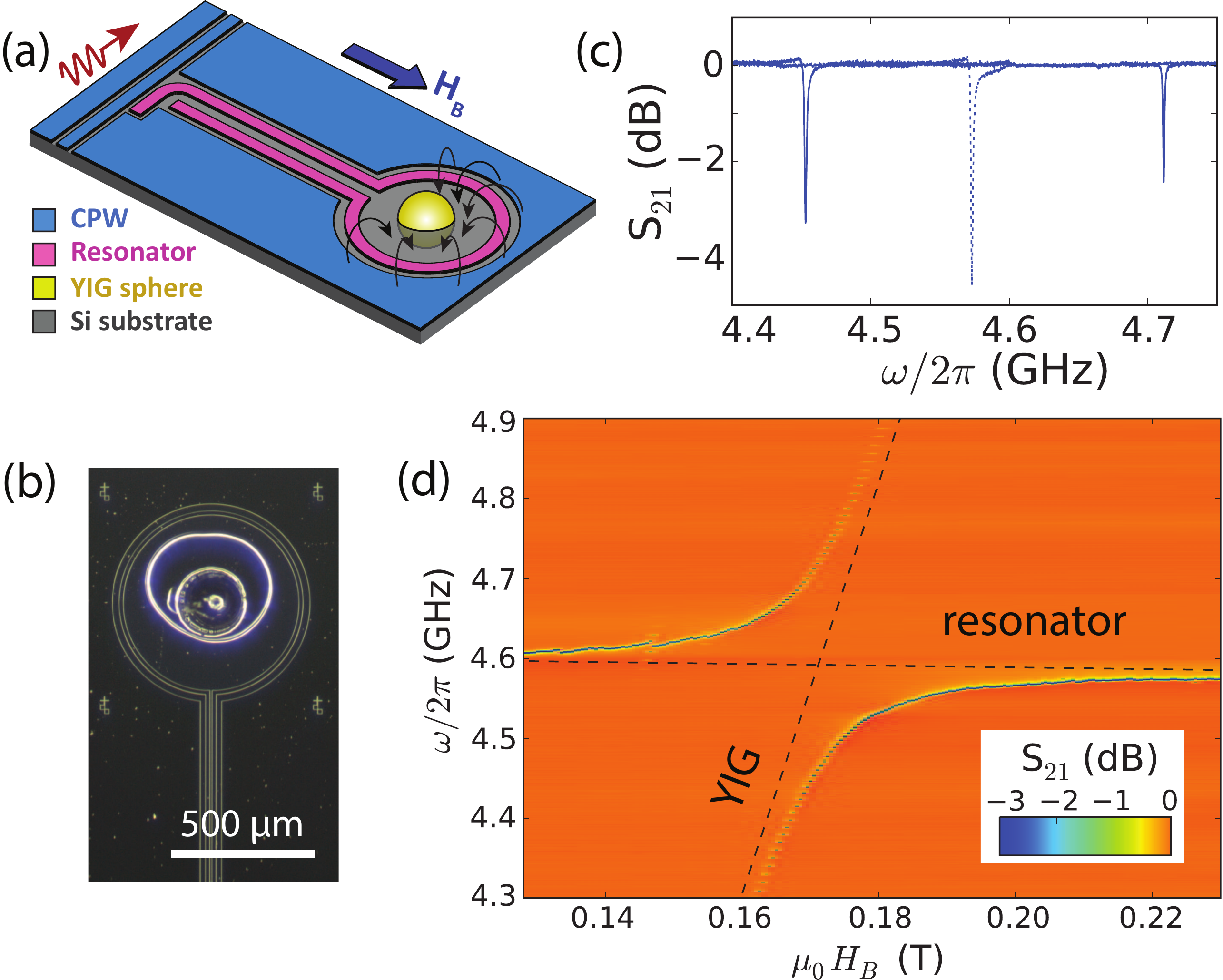}
 \caption{(a) Schematic of the one-YIG-sphere superconducting circuit design. (b) Microscope image of the circular antenna with a mounted YIG sphere. (c) Power transmission of the CPW feed line for the circular-antenna superconducting resonator. Solid spectrum: Rabi splitting ($\omega_c = \omega_m$) measured at $\mu_0H_B=0.171$ T. Dashed spectrum: uncoupled resonator photon mode ($|\omega_c - \omega_m| \gg g_c$) measured at $\mu_0H_B=0.25$ T. (d) Full power spectrum of the hybrid circuit showing magnon-photon mode repulsion.}
 \label{fig1}
\end{figure}

The microwave characterizations were conducted at 1.6 K using a vector network analyzer with an input power of -50 dBm \cite{LiPRL2019_magnon}. Fig. 1(d) shows the power transmission spectra of the resonator absorption. By sweeping the in-plane biasing field, mode anti-crossing with sharp lines is observed between the magnon mode of the YIG sphere ($\omega_m$) and photon mode of the superconducting resonator ($\omega_c$). When their frequencies are degenerate, large mode splitting between the two modes is observed (Fig. 1c). The asymmetric lineshapes are likely due to the complex impedance introduced by the resonator. The extracted magnon-photon coupling strength is $g_c/2\pi=130$ MHz; see the Supplemental Materials for more information for the fitting \cite{supplement}. For the 250-$\mu$m-diameter YIG sphere, the total spin number is $N=M_sV/\gamma(\hbar/2)$, where $M_s$ is the magnetization of YIG,  $V$ is the volume, and $\gamma/2\pi=28$ GHz/T is the gyromagnetic ratio. Calculation gives $N=1.2\times 10^{17}$. This yields a single-Bohr-magneton coupling strength of  $g_{c0}/2\pi=g_c/2\pi\sqrt{N}=0.38$ Hz, which is an order of magnitude larger than achieved in macroscopic cavities \cite{TabuchiPRL2014,ZhangPRL2014}. The intrinsic magnon and photon damping rates (full width at half maximum) are $\kappa_m/2\pi=1.0$ MHz and $\kappa_c/2\pi=1.3$ MHz, respectively, yielding a cooperativity of $C=g_c^2/\kappa_m\kappa_c=13000$. We do not observe any significant degradation of the superconducting resonator quality factor with the YIG sphere mounting, since neither the sphere nor mounting glue interfere with the impedance of the loop antenna.

It is worth noting that the circular design in Fig. 1(a) can efficiently suppress the crosstalk with nonuniform magnon modes and the associated decoherence process. This is important because any undesired mode coupling to the hybrid microwave circuit will cause additional non-resonant energy dissipation in the dispersive regime.

\begin{figure}[htb]
 \centering
 \includegraphics[width=3.0 in]{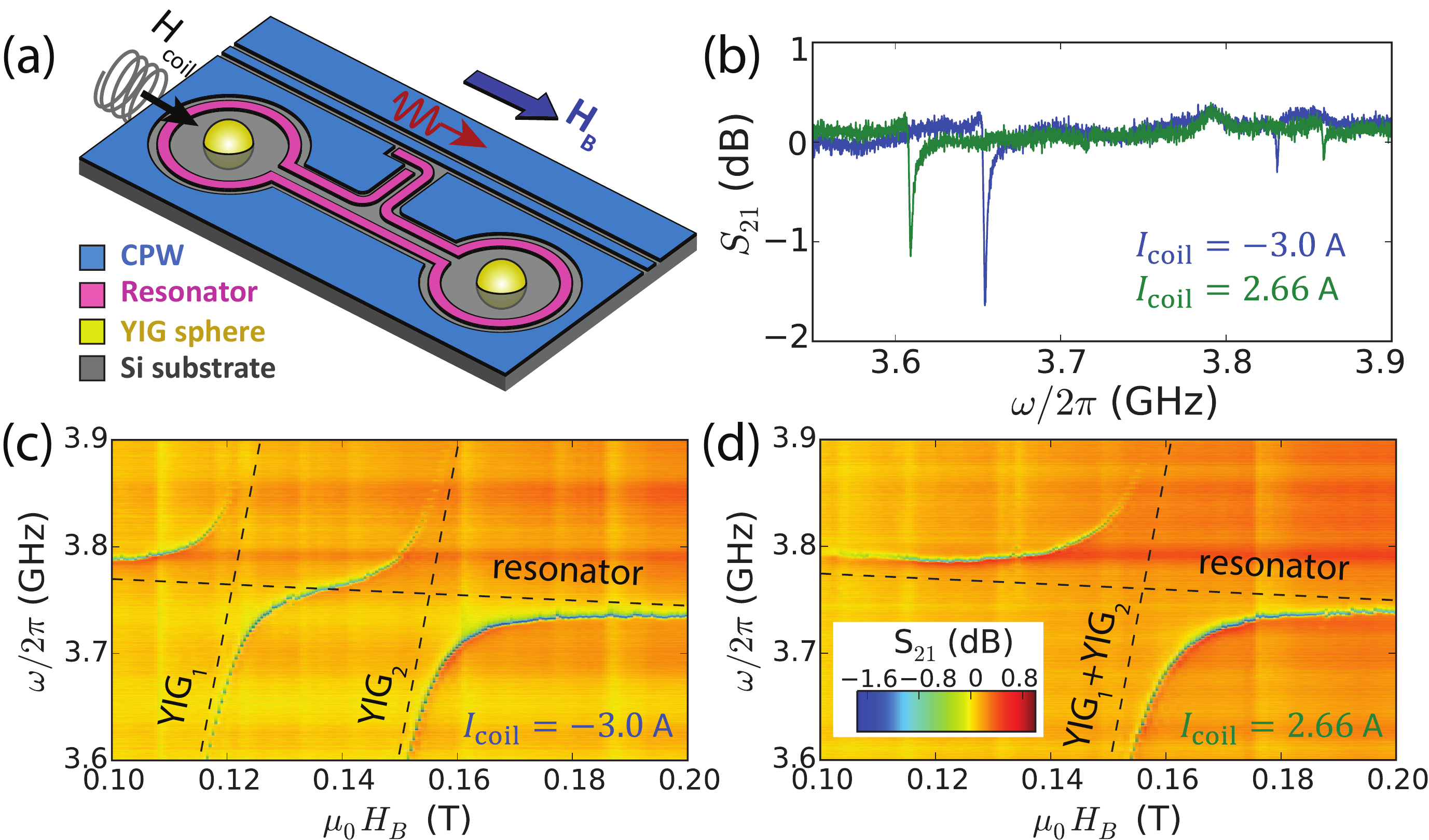}
 \caption{(a) Schematic of the two-YIG-sphere superconducting circuit design with an additional local NbTi superconducting coil. (b) Power transmission of the CPW feed line showing Rabi splitting of the resonator in couple with only one YIG sphere or both YIG spheres, measured at $\mu_0H_B=0.154$ T. Blue curve: $\omega_c = \omega_{m2}$ with $I_{coil}=-3.0$ A, where $\omega_{m1}$ is far detuned. Green curve: $\omega_c = \omega_{m1} = \omega_{m2}$ with $I_{coil}=2.66$ A. (c-d) Full power spectra of the hybrid magnonic circuit for (c) $I_{coil}=-3.0$ A and (d) $I_{coil}=2.66$ A.}
 \label{fig2}
\end{figure}

Next, we use the same circuit schematic to couple two remote YIG spheres. Shown in Fig. 2(a), we have extended the resonator design to include two circular antennas which are located symmetrically on the two sides of the resonator. Two YIG spheres are mounted at the centers of the antennas. The spheres are separated by 7 mm, which is 28 times of their diameters, and thus the magnetic dipolar interaction is suppressed. Electromagnetic simulation shows that the observed resonant frequency range (3.7-3.8 GHz) corresponds to a mode with both circular loops at the current antinode, allowing for maximal magnon-photon coupling efficiency (see the Supplemental Materials for details \cite{supplement}).

To allow for individual magnon frequency control, we have also integrated a local NbTi superconducting coil adjacent to one YIG sphere. The NbTi coil can generate a few hundred Oersted field onto the nearby YIG sphere without Ohmic heating in the cryogenic environment. Figs. 2(c-d) show the power transmission spectra of the circuit at two different coil currents ($I_{coil}$). At $I_{coil}=-3.0$ A, the magnon modes of the two YIG spheres are well detuned, leading to the observation of two separated avoided crossings with the superconducting resonator mode. The extracted magnon-photon coupling strengths are $g_{c1}/2\pi=81$ MHz and $g_{c2}/2\pi=88$ MHz. At $I_{coil}=2.66$ A where the two magnon modes are fully degenerate, a single avoided crossing is observed, with an stronger coupling strength of $g_{c1+2}/2\pi=121$ MHz. Comparing the two cases, a clear difference of the mode splitting lineshapes is observed at magnon-photon degeneracy (Fig. 2b), which confirms the change of magnon-photon coupling strength by coupling with two spheres instead of one.

The magnon-photon coupling strength can be expressed as \cite{LiJAP2020}:
\begin{equation}\label{eq01}
g_c = \sqrt{\omega_c\omega_MV_M \over 4V_c}
\end{equation}
where $\omega_M=\gamma\mu_0M_s$, and $V_M/V_c$ is the effective volume ratio between the YIG sphere and the resonator for providing inductance. When the two magnon frequencies are degenerate, their total volumes $V_M$ add up. Thus, the expected mutual coupling strength is $\sqrt{g_{c1}^2+g_{c2}^2 }/2\pi=120$ MHz, agreeing well with the measured $g_{c1+2}$. Comparing with the one-sphere circuit, since both circuits have the same antenna geometry design, same value of $V_M/V_c$ can be used for both circuits and the only difference is $\omega_c$. From Eq. (1), the coupling strength of the two-sphere circuit can be expressed as $g_{c1+2}^{theo}=g_c \sqrt{\omega_{c,\text{2YIG}}/\omega_{c,\text{1YIG}}}$. Taking $\omega_{c,\text{1YIG}}/2\pi=4.6$ GHz and $\omega_{c,\text{2YIG}}/2\pi=3.75$ GHz from Figs. 1 and 2, we obtain $g_{c1+2}^{theo}/2\pi=117$ MHz, again close to the experimental $g_{c1+2}$.

\begin{figure}[htb]
 \centering
 \includegraphics[width=3.0 in]{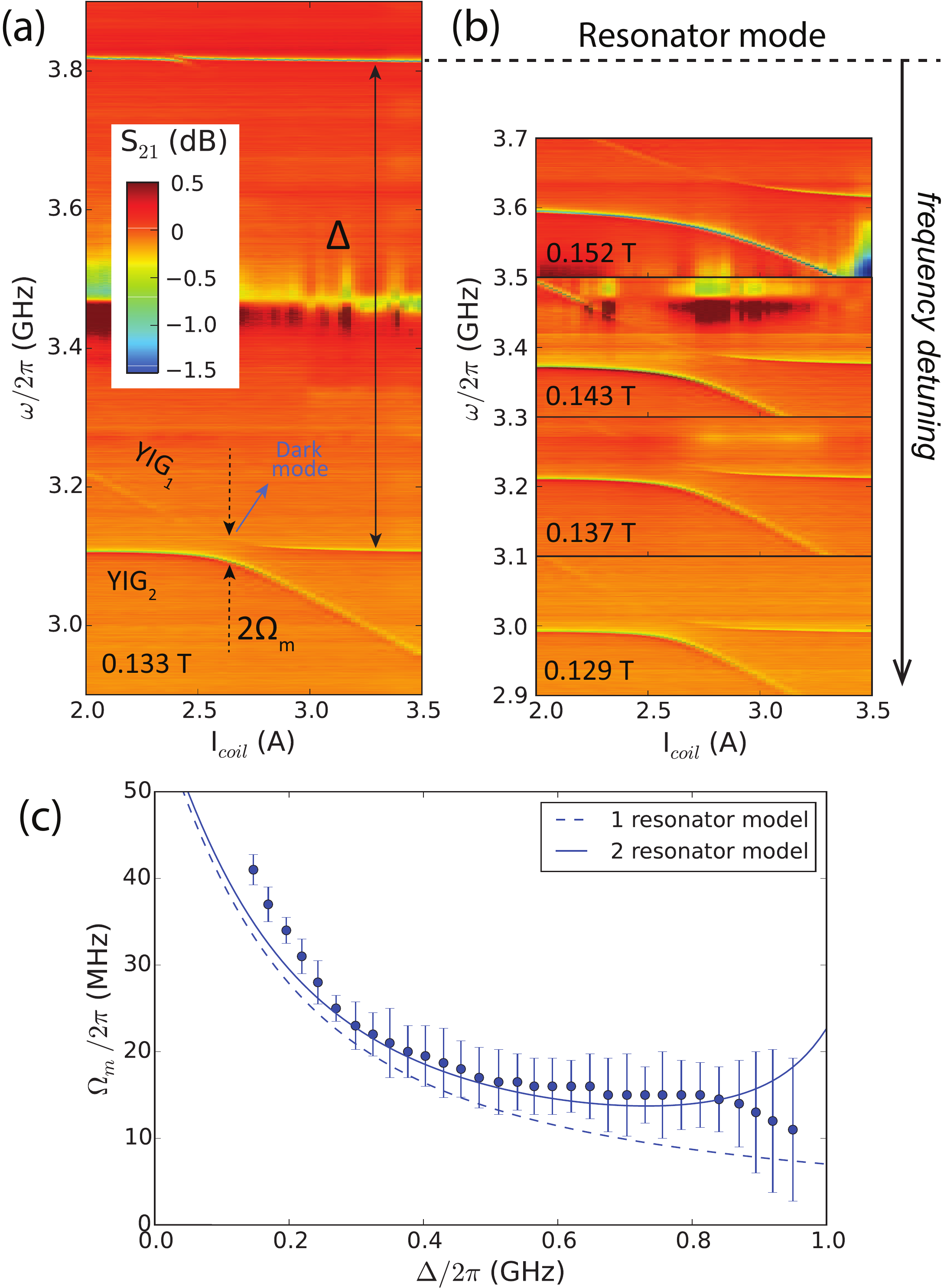}
 \caption{(a) Power spectra showing remote magnon-magnon coupling ($2\Omega_m$) in the dispersive regime, where $\Delta/2\pi=0.713$ GHz is much larger than $g_{c1}$, $g_{c2}$. Data are measured at $\mu_0H_B=0.133$ T. The gray arrow labels the dark mode. (b) Power spectra taken at different $\mu_0H_B$ showing the evolution of $\Omega_m$. (c) Extracted $\Omega_m$ as a function of $\Delta$. The error bars indicate single standard deviation uncertainties that arise primarily from the fitting of the resonances. Dashed curve: theoretic prediction by Eq. (2). Solid curve: adding an adjacent resonance mode which also contributes to dispersive magnon-magnon coupling \cite{supplement}.}
 \label{fig3}
\end{figure}

As a key achievement, we show that the coplanar superconducting circuit can enable remote coupling of the magnon modes between two distant YIG spheres. The coupling is achieved in the dispersive regime \cite{LambertPRA2016,RameshtiPRB2018,GrigoryanPRB2019,XuPRB2019}, i.e. where the resonator mode is well-detuned from the magnon frequencies (Fig. 3a). This regime has been routinely applied in quantum information processing in order to minimize qubit decoherence from being coupled to the external microwave circuits, including the recent demonstration of single-magnon operations with qubits \cite{LachanceQuirionScience2020}. Shown in Fig. 3(a), we fix the global biasing field to $\mu_0H_B=0.133$ T, so that the frequency detuning between the resonator mode and the magnon mode of YIG 2, $\Delta=|\omega_c-\omega_{m2}|$, is much larger than the magnon-resonator coupling strengths ($\Delta \gg g_{c1}, g_{c2}$). By ramping $I_{coil}$ to modify $\omega_{m1}$, a clear avoided crossing is observed between the magnon modes of the two YIG spheres, indicating their coherent coupling. The coupling strength is $\Omega_m/2\pi=16$ MHz, which is an order of magnitude larger than the magnon damping rates ($\kappa_{m1}/2\pi=1.8$ MHz, $\kappa_{m2}/2\pi=1.6$ MHz), leading to strong magnon-magnon coupling with a cooperativity of $C_m=\Omega_m^2/\kappa_{m1}\kappa_{m2}=89$. We also observe the dark mode of magnon-magnon coupling, which corresponds to the out-of-phase excitation of the two YIG spheres. It is decoupled from the superconducting resonator and thus cannot be excited and detected. This agrees with prior reports on similar hybrid magnonic systems \cite{LambertPRA2016,GrigoryanPRB2019,XuPRB2019,ZhangNComm2015}.

By repeating the measurement at different biasing fields to modify $\Delta$ (Fig. 3b), the extracted $\Omega_m$-$\Delta$ dependence shows an inversely proportional trend, which is plotted in Fig. 3(c). This trend can be captured by the theoretical model of cavity-mediated magnon-magnon coupling in the dispersive limit \cite{GrigoryanPRB2019,XuPRB2019}:
\begin{equation}\label{eq02}
\Omega_m = {g_{c1}g_{c2} \over \Delta}
\end{equation}
Taking the experimentally determined $g_{c1}$, $g_{c2}$ from the last section, the prediction from Eq. (2) (dashed blue curve in Fig. 3c) is close to the experimental results. \textcolor{black}{A small offset is likely due to additional channels of dispersive magnon-magnon coupling from other resonator modes. For example, by considering a mode at 2.76 GHz found in simulation (see the Supplemental Materials for details \cite{supplement}), this offset can be possibly reproduced by the revised 2-resonator model (solid blue curve).} The demonstration of nonlocal magnon-magnon coupling with a superconducting resonator is a critical step towards building distributed quantum magnonic networks such as generating entanglement over distances and implementing remote quantum memories.

\begin{figure}[htb]
 \centering
 \includegraphics[width=3.0 in]{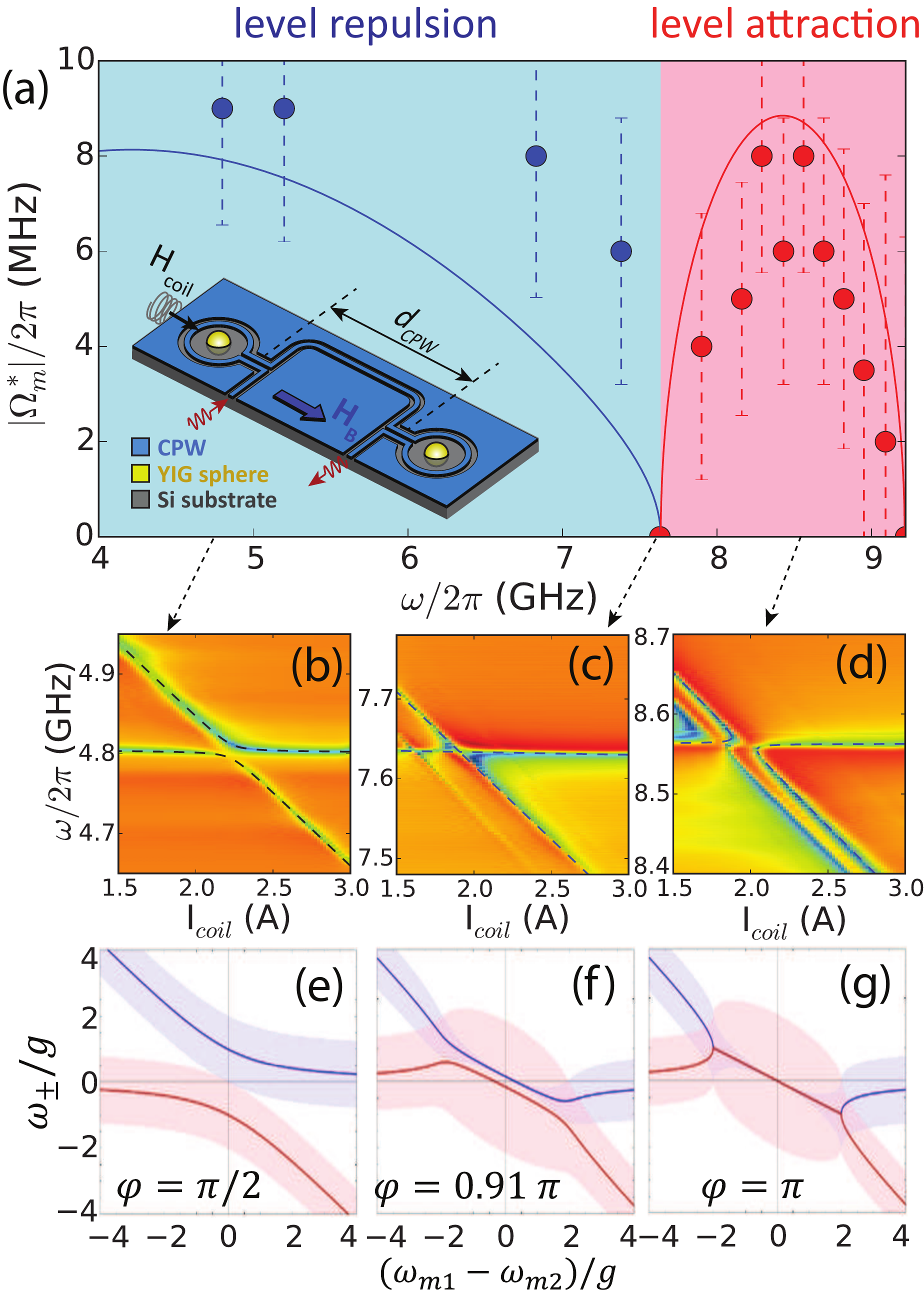}
 \caption{(a) Extracted magnon-magnon coupling strength $\Omega_m^*$ at different mode degenerate frequency. The error bars indicate single standard deviation uncertainties that arise primarily from the fitting of the resonances. Level repulsion and attraction are measured below and above $\omega/2\pi=7.63$ GHz, respectively. Inset: Schematic of the coplanar waveguide design for magnon-magnon level attraction. Solid curve: theoretical prediction using $\omega_0/2\pi=8.425$ GHz, $g_r/2\pi=8.5$ MHz, and $\epsilon = 0.29$. (b-d) Power spectra taken at (b) $\mu_0H_B=0.145$ T, (c) $\mu_0H_B=0.25$ T and (d) $\mu_0H_B=0.285$ T, corresponding to $\omega/2\pi=4.8$ GHz, 7.63 GHz and 8.56 GHz in (a). Dashed curves are fitting to Eq. (3) in (b) and (d) and a guide to eye in (c). (e-g) Theoretical prediction of magnon-magnon coupling mediated by propagating microwave with different $\varphi$ matching the conditions in (b-d).}
 \label{fig4}
\end{figure}

Lastly, we demonstrate that the hybrid circuit architecture can also implement level attraction of magnon modes between two YIG spheres, which has recently attracted wide interest for studying non-Hermitian physics and topological information processing \cite{HarderPRL2018,BhoiPRB2019,BoventerPRR2020}. Shown in Fig. 4(a) inset, the circuit consists of a coplanar waveguide with the signal line forming two circular loops where two YIG spheres are mounted at the center. The propagating microwave enables coherent magnon-magnon coupling between the two remote spheres due to their concerted absorption of microwave photons \cite{GrigoryanPRB2019,XuPRB2019}. At different intersection frequency, the two magnon modes show level repulsion (Fig. 4b), level attraction (Fig. 4d) and the transition state (Fig. 4c), which depends on the phase difference of propagating microwave between the two circular antennas as determined by the wavelength and frequency. The parasite diagonal modes are nonuniform magnon excitations due to impedance interruption during YIG sphere mounting \cite{supplement}.

The extracted magnon-magnon coupling strength $\Omega_m^*$ is summarized in Fig. 4(a) as a function of the intersection frequency. For the level attraction regime, we use a imaginary value of $\Omega_m^*$ as a fit parameter to the coupling spectrum and plot $|\Omega_m^*|$ instead. The regime of level attraction starts from 7.63 GHz, with the value of $|\Omega_m^*|$ reaching a plateau of 8 MHz centered at 8.4 GHz, then dropping again to zero at 9.2 GHz and switching back to level repulsion. This regime has been identified as dissipative coupling mediated by propagating microwave fields \cite{GrigoryanPRB2019,XuPRB2019}, with the coupling strength sensitively dependent on the phase offset $\varphi=(\omega/c_{eff})d_{CPW}$. Here $c_{eff}=c/\sqrt{\epsilon_{eff}}$ and the effective dielectric constant $\epsilon_{eff}=(\epsilon_{Si}+1)/2=6.34$ with $\epsilon_{Si}=11.68$. In order to reproduce the experimental observation, we have expanded the microwave transmission theory to account for the evolution of coupling with $\varphi$. The magnon resonance is introduced as a perturbation of the transmission matrix of the circular antenna in terms of frequency-dependent magnetic susceptibility. The magnon-magnon coupling is mediated by their mutual coupling to propagating microwave with a magnon-radiative coupling strength $g_r$ \cite{YaoPRB2019}; this term is also assumed to dominate the magnon damping rate over the intrinsic damping. In the limit of weak impedance mismatch from the circular antenna, the complex eigen-frequency of the collective dynamics of the two spheres can be expressed as:
\begin{equation}\label{eq03}
\omega_\pm = {\omega_{m1}+\omega_{m2} \over 2} - ig_r \pm \sqrt{\left(\omega_{m1}-\omega_{m2} \over 2\right)^2-g^2_re^{2i\varphi}}
\end{equation}
Eq. (3) yields a real coupling of $\Omega_m^*=g_r$ for $\varphi = \pi/2$ and an imaginary coupling of $\Omega_m^*=ig_r$ for $\varphi = \pi$. With the value of $d_{CPW}$, the location of $\varphi = \pi$ corresponds to a microwave frequency of $\omega/2\pi=8.3$ GHz, which matches with the plateau center of the dissipative coupling in Fig. 4(a).

To explain the transition from level repulsion to level attraction, we show the evolution of Eq. (3) in Figs. 4(e-g) at $\varphi = \pi/2$, $0.91\pi$ and $\pi$, with the real parts shown in curves and the imaginary parts shown in shaded areas. As $\varphi$ slightly deviates from $\pi$, the two collective frequencies move parallel to each other and at a very close distance in a wide interval of detuning. Due to a small frequency separation between the modes, the resonance linewidths of the modes overlap. This makes experimental distinguishing of both modes a difficult task, and leads to an observation of blurred level attraction with a reduced effective $\Omega_m^*$. \textcolor{black}{One example is $\omega/2\pi=7.63$ GHz in Fig. 4(a), where the blurred transmission spectra in Fig. 4(c) cannot be fitted and a coupling of $\Omega_m/2\pi=0$ is deduced without an errorbar (same for $\omega/2\pi=9.22$ GHz).} As a simple criterion of mode distinguishability, we assume that the modes become indistinguishable when $|\omega_+-\omega_-|<\epsilon g_r$, where $\epsilon$ is a dimensionless parameter; see the Supplemental Materials for detailed discussion \cite{supplement}. Following this approach, we find that the onset of level attraction happens when $|\sin{\varphi} |<\epsilon$. Fig. 4 shows the theoretical prediction of $\Omega_m^*$ with the optimized fit parameters $\epsilon=0.29$ and $g_r/2\pi=8.5$ MHz that yield the least variance, which reasonably agrees with the experimental data. Thus, the hybrid magnonics circuit platform can be used to implement and control the coupling phenomena, from coherent coupling to dissipative coupling, with the coupling parameters highly adjustable by the circuit geometry and design.

The successful integration of hybrid magnonics into a low-loss superconducting circuit, with demonstrations of coherent and dissipative couplings between two remote YIG spheres, provides a circuit platform for building prototype quantum magnonic network \cite{ZhangNComm2015,YuPRB2020}. This compact circuit architecture has the potential for implementing desirable quantum operations with magnons, including coherent microwave-to-optical transduction, nonreciprocity, and magnon-enabled dark matter sensing \cite{TricklePRL2020}. It is worth noting that YIG thin-film magnonic devices can undergo fine nanofabrications for wafer-scale integration \cite{HeinzNanoLett2020}, and GGG-free high-quality YIG thin film growth \cite{FanPRApplied2020} may provide an ultimate solution for scalable hybrid magnonic network. Finally, we also point out that the Si-based circuit design can incorporate Si integrated photonics for further empowering the quantum network with magnons, microwave and optics.

All work at Argonne was supported by the U.S. Department of Energy, Office of Science, Basic Energy Sciences, Materials Sciences and Engineering Division. Use of the Center for Nanoscale Materials (CNM), an Office of Science user facility, was supported by the U.S. Department of Energy, Office of Science, Office of Basic Energy Sciences, under Contract no. DE-AC02-06CH11357. The contributions from V.G.Y., M.L., T.W.C, P.B, to the superconducting resonator design and Si etching was supported by the U.S. Department of Energy, Office of Science, High-Energy Physics. The work by C.T. and V.T. on the theoretical analysis of the experimental data was supported by the DARPA TWEED grant DARPA-PA19-04-05-FP-001 and by the Oakland University Foundation.


\begin{thebibliography}{53}%
\makeatletter
\providecommand \@ifxundefined [1]{%
 \@ifx{#1\undefined}
}%
\providecommand \@ifnum [1]{%
 \ifnum #1\expandafter \@firstoftwo
 \else \expandafter \@secondoftwo
 \fi
}%
\providecommand \@ifx [1]{%
 \ifx #1\expandafter \@firstoftwo
 \else \expandafter \@secondoftwo
 \fi
}%
\providecommand \natexlab [1]{#1}%
\providecommand \enquote  [1]{``#1''}%
\providecommand \bibnamefont  [1]{#1}%
\providecommand \bibfnamefont [1]{#1}%
\providecommand \citenamefont [1]{#1}%
\providecommand \href@noop [0]{\@secondoftwo}%
\providecommand \href [0]{\begingroup \@sanitize@url \@href}%
\providecommand \@href[1]{\@@startlink{#1}\@@href}%
\providecommand \@@href[1]{\endgroup#1\@@endlink}%
\providecommand \@sanitize@url [0]{\catcode `\\12\catcode `\$12\catcode
  `\&12\catcode `\#12\catcode `\^12\catcode `\_12\catcode `\%12\relax}%
\providecommand \@@startlink[1]{}%
\providecommand \@@endlink[0]{}%
\providecommand \url  [0]{\begingroup\@sanitize@url \@url }%
\providecommand \@url [1]{\endgroup\@href {#1}{\urlprefix }}%
\providecommand \urlprefix  [0]{URL }%
\providecommand \Eprint [0]{\href }%
\providecommand \doibase [0]{http://dx.doi.org/}%
\providecommand \selectlanguage [0]{\@gobble}%
\providecommand \bibinfo  [0]{\@secondoftwo}%
\providecommand \bibfield  [0]{\@secondoftwo}%
\providecommand \translation [1]{[#1]}%
\providecommand \BibitemOpen [0]{}%
\providecommand \bibitemStop [0]{}%
\providecommand \bibitemNoStop [0]{.\EOS\space}%
\providecommand \EOS [0]{\spacefactor3000\relax}%
\providecommand \BibitemShut  [1]{\csname bibitem#1\endcsname}%
\let\auto@bib@innerbib\@empty
\bibitem [{\citenamefont {Kurizki}\ \emph {et~al.}(2015)\citenamefont
  {Kurizki}, \citenamefont {Bertet}, \citenamefont {Kubo}, \citenamefont
  {M{\o}lmer}, \citenamefont {Petrosyan}, \citenamefont {Rabl},\ and\
  \citenamefont {Schmiedmayer}}]{KurizkiPNAS2015}%
  \BibitemOpen
  \bibfield  {author} {\bibinfo {author} {\bibfnamefont {G.}~\bibnamefont
  {Kurizki}}, \bibinfo {author} {\bibfnamefont {P.}~\bibnamefont {Bertet}},
  \bibinfo {author} {\bibfnamefont {Y.}~\bibnamefont {Kubo}}, \bibinfo {author}
  {\bibfnamefont {K.}~\bibnamefont {M{\o}lmer}}, \bibinfo {author}
  {\bibfnamefont {D.}~\bibnamefont {Petrosyan}}, \bibinfo {author}
  {\bibfnamefont {P.}~\bibnamefont {Rabl}}, \ and\ \bibinfo {author}
  {\bibfnamefont {J.}~\bibnamefont {Schmiedmayer}},\ }\href {\doibase
  10.1073/pnas.1419326112} {\bibfield  {journal} {\bibinfo  {journal} {Proc.
  Natl. Acad. Sci.}\ }\textbf {\bibinfo {volume} {112}},\ \bibinfo {pages}
  {3866} (\bibinfo {year} {2015})}\BibitemShut {NoStop}%
\bibitem [{\citenamefont {Clerk}\ \emph {et~al.}(2020)\citenamefont {Clerk},
  \citenamefont {Lehnert}, \citenamefont {Bertet}, \citenamefont {Petta},\ and\
  \citenamefont {Nakamura}}]{ClerkNPhys2020}%
  \BibitemOpen
  \bibfield  {author} {\bibinfo {author} {\bibfnamefont {A.~A.}\ \bibnamefont
  {Clerk}}, \bibinfo {author} {\bibfnamefont {K.~W.}\ \bibnamefont {Lehnert}},
  \bibinfo {author} {\bibfnamefont {P.}~\bibnamefont {Bertet}}, \bibinfo
  {author} {\bibfnamefont {J.~R.}\ \bibnamefont {Petta}}, \ and\ \bibinfo
  {author} {\bibfnamefont {Y.}~\bibnamefont {Nakamura}},\ }\href@noop {}
  {\bibfield  {journal} {\bibinfo  {journal} {Nature Phys.}\ }\textbf {\bibinfo
  {volume} {16}},\ \bibinfo {pages} {257} (\bibinfo {year} {2020})}\BibitemShut
  {NoStop}%
\bibitem [{\citenamefont {Wallraff}\ \emph {et~al.}(2004)\citenamefont
  {Wallraff}, \citenamefont {Schuster}, \citenamefont {Blais}, \citenamefont
  {Frunzio}, \citenamefont {Huang}, \citenamefont {Majer}, \citenamefont
  {Kumar},\ and\ \citenamefont {Schoelkopf}}]{WallraffNature2004}%
  \BibitemOpen
  \bibfield  {author} {\bibinfo {author} {\bibfnamefont {A.}~\bibnamefont
  {Wallraff}}, \bibinfo {author} {\bibfnamefont {D.~I.}\ \bibnamefont
  {Schuster}}, \bibinfo {author} {\bibfnamefont {A.}~\bibnamefont {Blais}},
  \bibinfo {author} {\bibfnamefont {L.}~\bibnamefont {Frunzio}}, \bibinfo
  {author} {\bibfnamefont {R.-S.}\ \bibnamefont {Huang}}, \bibinfo {author}
  {\bibfnamefont {J.}~\bibnamefont {Majer}}, \bibinfo {author} {\bibfnamefont
  {S.~M.}\ \bibnamefont {Kumar}, \bibfnamefont {S.~abd~Girvin}}, \ and\
  \bibinfo {author} {\bibfnamefont {R.~J.}\ \bibnamefont {Schoelkopf}},\
  }\href@noop {} {\bibfield  {journal} {\bibinfo  {journal} {Nature}\ }\textbf
  {\bibinfo {volume} {431}},\ \bibinfo {pages} {162} (\bibinfo {year}
  {2004})}\BibitemShut {NoStop}%
\bibitem [{\citenamefont {Zhu}\ \emph {et~al.}(2018)\citenamefont {Zhu},
  \citenamefont {Zhao}, \citenamefont {Choi}, \citenamefont {Lu}, \citenamefont
  {Dane}, \citenamefont {Englund},\ and\ \citenamefont
  {Berggren}}]{ZhuNatureNano2018}%
  \BibitemOpen
  \bibfield  {author} {\bibinfo {author} {\bibfnamefont {D.}~\bibnamefont
  {Zhu}}, \bibinfo {author} {\bibfnamefont {Q.-Y.}\ \bibnamefont {Zhao}},
  \bibinfo {author} {\bibfnamefont {H.}~\bibnamefont {Choi}}, \bibinfo {author}
  {\bibfnamefont {T.-J.}\ \bibnamefont {Lu}}, \bibinfo {author} {\bibfnamefont
  {A.~E.}\ \bibnamefont {Dane}}, \bibinfo {author} {\bibfnamefont
  {D.}~\bibnamefont {Englund}}, \ and\ \bibinfo {author} {\bibfnamefont
  {K.~K.}\ \bibnamefont {Berggren}},\ }\href@noop {} {\bibfield  {journal}
  {\bibinfo  {journal} {Nature Nanotech.}\ }\textbf {\bibinfo {volume} {13}},\
  \bibinfo {pages} {506} (\bibinfo {year} {2018})}\BibitemShut {NoStop}%
\bibitem [{\citenamefont {Fan}\ \emph {et~al.}(2018)\citenamefont {Fan},
  \citenamefont {Zou}, \citenamefont {Cheng}, \citenamefont {Guo},
  \citenamefont {Han}, \citenamefont {Gong}, \citenamefont {Wang},\ and\
  \citenamefont {Tang}}]{FanSciAdv2018}%
  \BibitemOpen
  \bibfield  {author} {\bibinfo {author} {\bibfnamefont {L.}~\bibnamefont
  {Fan}}, \bibinfo {author} {\bibfnamefont {C.-L.}\ \bibnamefont {Zou}},
  \bibinfo {author} {\bibfnamefont {R.}~\bibnamefont {Cheng}}, \bibinfo
  {author} {\bibfnamefont {X.}~\bibnamefont {Guo}}, \bibinfo {author}
  {\bibfnamefont {X.}~\bibnamefont {Han}}, \bibinfo {author} {\bibfnamefont
  {Z.}~\bibnamefont {Gong}}, \bibinfo {author} {\bibfnamefont {S.}~\bibnamefont
  {Wang}}, \ and\ \bibinfo {author} {\bibfnamefont {H.~X.}\ \bibnamefont
  {Tang}},\ }\href@noop {} {\bibfield  {journal} {\bibinfo  {journal} {Science
  Adv.}\ }\textbf {\bibinfo {volume} {4}},\ \bibinfo {pages} {eaar4994}
  (\bibinfo {year} {2018})}\BibitemShut {NoStop}%
\bibitem [{\citenamefont {Chu}\ \emph {et~al.}(2017)\citenamefont {Chu},
  \citenamefont {Kharel}, \citenamefont {Renninger}, \citenamefont {Burkhart},
  \citenamefont {Frunzio}, \citenamefont {Rakich},\ and\ \citenamefont
  {Schoelkopf}}]{ChuScience2017}%
  \BibitemOpen
  \bibfield  {author} {\bibinfo {author} {\bibfnamefont {Y.}~\bibnamefont
  {Chu}}, \bibinfo {author} {\bibfnamefont {P.}~\bibnamefont {Kharel}},
  \bibinfo {author} {\bibfnamefont {W.~H.}\ \bibnamefont {Renninger}}, \bibinfo
  {author} {\bibfnamefont {L.~D.}\ \bibnamefont {Burkhart}}, \bibinfo {author}
  {\bibfnamefont {L.}~\bibnamefont {Frunzio}}, \bibinfo {author} {\bibfnamefont
  {P.~T.}\ \bibnamefont {Rakich}}, \ and\ \bibinfo {author} {\bibfnamefont
  {R.~J.}\ \bibnamefont {Schoelkopf}},\ }\href@noop {} {\bibfield  {journal}
  {\bibinfo  {journal} {Science}\ }\textbf {\bibinfo {volume} {358}},\ \bibinfo
  {pages} {199} (\bibinfo {year} {2017})}\BibitemShut {NoStop}%
\bibitem [{\citenamefont {Satzinger}\ \emph {et~al.}(2018)\citenamefont
  {Satzinger}, \citenamefont {Zhong}, \citenamefont {Chang}, \citenamefont
  {Peairs}, \citenamefont {Bienfait}, \citenamefont {Chou}, \citenamefont
  {Cleland}, \citenamefont {Conner}, \citenamefont {Dumur}, \citenamefont
  {Grebel}, \citenamefont {Gutierrez}, \citenamefont {November}, \citenamefont
  {Povey}, \citenamefont {Whiteley}, \citenamefont {Awschalom}, \citenamefont
  {Schuster},\ and\ \citenamefont {Cleland}}]{SatzingerNature2018}%
  \BibitemOpen
  \bibfield  {author} {\bibinfo {author} {\bibfnamefont {K.~J.}\ \bibnamefont
  {Satzinger}}, \bibinfo {author} {\bibfnamefont {Y.~P.}\ \bibnamefont
  {Zhong}}, \bibinfo {author} {\bibfnamefont {H.-S.}\ \bibnamefont {Chang}},
  \bibinfo {author} {\bibfnamefont {G.~A.}\ \bibnamefont {Peairs}}, \bibinfo
  {author} {\bibfnamefont {A.}~\bibnamefont {Bienfait}}, \bibinfo {author}
  {\bibfnamefont {M.-H.}\ \bibnamefont {Chou}}, \bibinfo {author}
  {\bibfnamefont {A.~Y.}\ \bibnamefont {Cleland}}, \bibinfo {author}
  {\bibfnamefont {C.~R.}\ \bibnamefont {Conner}}, \bibinfo {author}
  {\bibfnamefont {E.}~\bibnamefont {Dumur}}, \bibinfo {author} {\bibfnamefont
  {J.}~\bibnamefont {Grebel}}, \bibinfo {author} {\bibfnamefont
  {I.}~\bibnamefont {Gutierrez}}, \bibinfo {author} {\bibfnamefont {B.~H.}\
  \bibnamefont {November}}, \bibinfo {author} {\bibfnamefont {R.~G.}\
  \bibnamefont {Povey}}, \bibinfo {author} {\bibfnamefont {S.~J.}\ \bibnamefont
  {Whiteley}}, \bibinfo {author} {\bibfnamefont {D.~D.}\ \bibnamefont
  {Awschalom}}, \bibinfo {author} {\bibfnamefont {D.~I.}\ \bibnamefont
  {Schuster}}, \ and\ \bibinfo {author} {\bibfnamefont {A.~N.}\ \bibnamefont
  {Cleland}},\ }\href@noop {} {\bibfield  {journal} {\bibinfo  {journal}
  {Nature}\ }\textbf {\bibinfo {volume} {563}},\ \bibinfo {pages} {661}
  (\bibinfo {year} {2018})}\BibitemShut {NoStop}%
\bibitem [{\citenamefont {Schuster}\ \emph {et~al.}(2010)\citenamefont
  {Schuster}, \citenamefont {Sears}, \citenamefont {Ginossar}, \citenamefont
  {DiCarlo}, \citenamefont {Frunzio}, \citenamefont {Morton}, \citenamefont
  {Wu}, \citenamefont {Briggs}, \citenamefont {Buckley}, \citenamefont
  {Awschalom},\ and\ \citenamefont {Schoelkopf}}]{SchusterPRL2010}%
  \BibitemOpen
  \bibfield  {author} {\bibinfo {author} {\bibfnamefont {D.~I.}\ \bibnamefont
  {Schuster}}, \bibinfo {author} {\bibfnamefont {A.~P.}\ \bibnamefont {Sears}},
  \bibinfo {author} {\bibfnamefont {E.}~\bibnamefont {Ginossar}}, \bibinfo
  {author} {\bibfnamefont {L.}~\bibnamefont {DiCarlo}}, \bibinfo {author}
  {\bibfnamefont {L.}~\bibnamefont {Frunzio}}, \bibinfo {author} {\bibfnamefont
  {J.~J.~L.}\ \bibnamefont {Morton}}, \bibinfo {author} {\bibfnamefont
  {H.}~\bibnamefont {Wu}}, \bibinfo {author} {\bibfnamefont {G.~A.~D.}\
  \bibnamefont {Briggs}}, \bibinfo {author} {\bibfnamefont {B.~B.}\
  \bibnamefont {Buckley}}, \bibinfo {author} {\bibfnamefont {D.~D.}\
  \bibnamefont {Awschalom}}, \ and\ \bibinfo {author} {\bibfnamefont {R.~J.}\
  \bibnamefont {Schoelkopf}},\ }\href {\doibase 10.1103/PhysRevLett.105.140501}
  {\bibfield  {journal} {\bibinfo  {journal} {Phys. Rev. Lett.}\ }\textbf
  {\bibinfo {volume} {105}},\ \bibinfo {pages} {140501} (\bibinfo {year}
  {2010})}\BibitemShut {NoStop}%
\bibitem [{\citenamefont {Kubo}\ \emph {et~al.}(2010)\citenamefont {Kubo},
  \citenamefont {Ong}, \citenamefont {Bertet}, \citenamefont {Vion},
  \citenamefont {Jacques}, \citenamefont {Zheng}, \citenamefont {Dr\'{e}au},
  \citenamefont {Roch}, \citenamefont {Auffeves}, \citenamefont {Jelezko},
  \citenamefont {Wrachtrup}, \citenamefont {Barthe}, \citenamefont {Bergonzo},\
  and\ \citenamefont {Esteve}}]{KuboPRL2010}%
  \BibitemOpen
  \bibfield  {author} {\bibinfo {author} {\bibfnamefont {Y.}~\bibnamefont
  {Kubo}}, \bibinfo {author} {\bibfnamefont {F.~R.}\ \bibnamefont {Ong}},
  \bibinfo {author} {\bibfnamefont {P.}~\bibnamefont {Bertet}}, \bibinfo
  {author} {\bibfnamefont {D.}~\bibnamefont {Vion}}, \bibinfo {author}
  {\bibfnamefont {V.}~\bibnamefont {Jacques}}, \bibinfo {author} {\bibfnamefont
  {D.}~\bibnamefont {Zheng}}, \bibinfo {author} {\bibfnamefont
  {A.}~\bibnamefont {Dr\'{e}au}}, \bibinfo {author} {\bibfnamefont {J.-F.}\
  \bibnamefont {Roch}}, \bibinfo {author} {\bibfnamefont {A.}~\bibnamefont
  {Auffeves}}, \bibinfo {author} {\bibfnamefont {F.}~\bibnamefont {Jelezko}},
  \bibinfo {author} {\bibfnamefont {J.}~\bibnamefont {Wrachtrup}}, \bibinfo
  {author} {\bibfnamefont {M.~F.}\ \bibnamefont {Barthe}}, \bibinfo {author}
  {\bibfnamefont {P.}~\bibnamefont {Bergonzo}}, \ and\ \bibinfo {author}
  {\bibfnamefont {D.}~\bibnamefont {Esteve}},\ }\href {\doibase
  10.1103/PhysRevLett.105.140502} {\bibfield  {journal} {\bibinfo  {journal}
  {Phys. Rev. Lett.}\ }\textbf {\bibinfo {volume} {105}},\ \bibinfo {pages}
  {140502} (\bibinfo {year} {2010})}\BibitemShut {NoStop}%
\bibitem [{\citenamefont {Li}\ \emph {et~al.}(2019)\citenamefont {Li},
  \citenamefont {Polakovic}, \citenamefont {Wang}, \citenamefont {Xu},
  \citenamefont {Lendinez}, \citenamefont {Zhang}, \citenamefont {Ding},
  \citenamefont {Khaire}, \citenamefont {Saglam}, \citenamefont {Divan},
  \citenamefont {Pearson}, \citenamefont {Kwok}, \citenamefont {Xiao},
  \citenamefont {Novosad}, \citenamefont {Hoffmann},\ and\ \citenamefont
  {Zhang}}]{LiPRL2019_magnon}%
  \BibitemOpen
  \bibfield  {author} {\bibinfo {author} {\bibfnamefont {Y.}~\bibnamefont
  {Li}}, \bibinfo {author} {\bibfnamefont {T.}~\bibnamefont {Polakovic}},
  \bibinfo {author} {\bibfnamefont {Y.-L.}\ \bibnamefont {Wang}}, \bibinfo
  {author} {\bibfnamefont {J.}~\bibnamefont {Xu}}, \bibinfo {author}
  {\bibfnamefont {S.}~\bibnamefont {Lendinez}}, \bibinfo {author}
  {\bibfnamefont {Z.}~\bibnamefont {Zhang}}, \bibinfo {author} {\bibfnamefont
  {J.}~\bibnamefont {Ding}}, \bibinfo {author} {\bibfnamefont {T.}~\bibnamefont
  {Khaire}}, \bibinfo {author} {\bibfnamefont {H.}~\bibnamefont {Saglam}},
  \bibinfo {author} {\bibfnamefont {R.}~\bibnamefont {Divan}}, \bibinfo
  {author} {\bibfnamefont {J.}~\bibnamefont {Pearson}}, \bibinfo {author}
  {\bibfnamefont {W.-K.}\ \bibnamefont {Kwok}}, \bibinfo {author}
  {\bibfnamefont {Z.}~\bibnamefont {Xiao}}, \bibinfo {author} {\bibfnamefont
  {V.}~\bibnamefont {Novosad}}, \bibinfo {author} {\bibfnamefont
  {A.}~\bibnamefont {Hoffmann}}, \ and\ \bibinfo {author} {\bibfnamefont
  {W.}~\bibnamefont {Zhang}},\ }\href {\doibase 10.1103/PhysRevLett.123.107701}
  {\bibfield  {journal} {\bibinfo  {journal} {Phys. Rev. Lett.}\ }\textbf
  {\bibinfo {volume} {123}},\ \bibinfo {pages} {107701} (\bibinfo {year}
  {2019})}\BibitemShut {NoStop}%
\bibitem [{\citenamefont {Hou}\ and\ \citenamefont {Liu}(2019)}]{HouPRL2019}%
  \BibitemOpen
  \bibfield  {author} {\bibinfo {author} {\bibfnamefont {J.~T.}\ \bibnamefont
  {Hou}}\ and\ \bibinfo {author} {\bibfnamefont {L.}~\bibnamefont {Liu}},\
  }\href {\doibase 10.1103/PhysRevLett.123.107702} {\bibfield  {journal}
  {\bibinfo  {journal} {Phys. Rev. Lett.}\ }\textbf {\bibinfo {volume} {123}},\
  \bibinfo {pages} {107702} (\bibinfo {year} {2019})}\BibitemShut {NoStop}%
\bibitem [{\citenamefont {McKenzie-Sell}\ \emph {et~al.}(2019)\citenamefont
  {McKenzie-Sell}, \citenamefont {Xie}, \citenamefont {Lee}, \citenamefont
  {Robinson}, \citenamefont {Ciccarelli},\ and\ \citenamefont
  {Haigh}}]{McKenziePRB2019}%
  \BibitemOpen
  \bibfield  {author} {\bibinfo {author} {\bibfnamefont {L.}~\bibnamefont
  {McKenzie-Sell}}, \bibinfo {author} {\bibfnamefont {J.}~\bibnamefont {Xie}},
  \bibinfo {author} {\bibfnamefont {C.-M.}\ \bibnamefont {Lee}}, \bibinfo
  {author} {\bibfnamefont {J.~W.~A.}\ \bibnamefont {Robinson}}, \bibinfo
  {author} {\bibfnamefont {C.}~\bibnamefont {Ciccarelli}}, \ and\ \bibinfo
  {author} {\bibfnamefont {J.~A.}\ \bibnamefont {Haigh}},\ }\href {\doibase
  10.1103/PhysRevB.99.140414} {\bibfield  {journal} {\bibinfo  {journal} {Phys.
  Rev. B}\ }\textbf {\bibinfo {volume} {99}},\ \bibinfo {pages} {140414}
  (\bibinfo {year} {2019})}\BibitemShut {NoStop}%
\bibitem [{\citenamefont {Lachance-Quirion}\ \emph {et~al.}(2019)\citenamefont
  {Lachance-Quirion}, \citenamefont {Tabuchi}, \citenamefont {Gloppe},
  \citenamefont {Usami},\ and\ \citenamefont
  {Nakamura}}]{LachanceQuirionAPEx2019}%
  \BibitemOpen
  \bibfield  {author} {\bibinfo {author} {\bibfnamefont {D.}~\bibnamefont
  {Lachance-Quirion}}, \bibinfo {author} {\bibfnamefont {Y.}~\bibnamefont
  {Tabuchi}}, \bibinfo {author} {\bibfnamefont {A.}~\bibnamefont {Gloppe}},
  \bibinfo {author} {\bibfnamefont {K.}~\bibnamefont {Usami}}, \ and\ \bibinfo
  {author} {\bibfnamefont {Y.}~\bibnamefont {Nakamura}},\ }\href@noop {}
  {\bibfield  {journal} {\bibinfo  {journal} {Appl. Phys. Express}\ }\textbf
  {\bibinfo {volume} {12}},\ \bibinfo {pages} {070101} (\bibinfo {year}
  {2019})}\BibitemShut {NoStop}%
\bibitem [{\citenamefont {Li}\ \emph {et~al.}(2020)\citenamefont {Li},
  \citenamefont {Zhang}, \citenamefont {Tyberkevych}, \citenamefont {Kwok},
  \citenamefont {Hoffmann},\ and\ \citenamefont {Novosad}}]{LiJAP2020}%
  \BibitemOpen
  \bibfield  {author} {\bibinfo {author} {\bibfnamefont {Y.}~\bibnamefont
  {Li}}, \bibinfo {author} {\bibfnamefont {W.}~\bibnamefont {Zhang}}, \bibinfo
  {author} {\bibfnamefont {V.}~\bibnamefont {Tyberkevych}}, \bibinfo {author}
  {\bibfnamefont {W.-K.}\ \bibnamefont {Kwok}}, \bibinfo {author}
  {\bibfnamefont {A.}~\bibnamefont {Hoffmann}}, \ and\ \bibinfo {author}
  {\bibfnamefont {V.}~\bibnamefont {Novosad}},\ }\href@noop {} {\bibfield
  {journal} {\bibinfo  {journal} {J. Appl. Phys.}\ }\textbf {\bibinfo {volume}
  {128}},\ \bibinfo {pages} {130902} (\bibinfo {year} {2020})}\BibitemShut
  {NoStop}%
\bibitem [{\citenamefont {Rameshti~et al.}()}]{RameshtiarXiv2021}%
  \BibitemOpen
  \bibfield  {author} {\bibinfo {author} {\bibfnamefont {B.~Z.}\ \bibnamefont
  {Rameshti~et al.}},\ }\href@noop {} {\bibfield  {journal} {\bibinfo
  {journal} {arXiv}\ }}\bibinfo {note} {2106.09312}\BibitemShut {NoStop}%
\bibitem [{\citenamefont {Soykal}\ and\ \citenamefont
  {Flatt\'e}(2010)}]{SoykalPRL2010}%
  \BibitemOpen
  \bibfield  {author} {\bibinfo {author} {\bibfnamefont {O.~O.}\ \bibnamefont
  {Soykal}}\ and\ \bibinfo {author} {\bibfnamefont {M.~E.}\ \bibnamefont
  {Flatt\'e}},\ }\href {\doibase 10.1103/PhysRevLett.104.077202} {\bibfield
  {journal} {\bibinfo  {journal} {Phys. Rev. Lett.}\ }\textbf {\bibinfo
  {volume} {104}},\ \bibinfo {pages} {077202} (\bibinfo {year}
  {2010})}\BibitemShut {NoStop}%
\bibitem [{\citenamefont {Huebl}\ \emph {et~al.}(2013)\citenamefont {Huebl},
  \citenamefont {Zollitsch}, \citenamefont {Lotze}, \citenamefont {Hocke},
  \citenamefont {Greifenstein}, \citenamefont {Marx}, \citenamefont {Gross},\
  and\ \citenamefont {Goennenwein}}]{HueblPRL2013}%
  \BibitemOpen
  \bibfield  {author} {\bibinfo {author} {\bibfnamefont {H.}~\bibnamefont
  {Huebl}}, \bibinfo {author} {\bibfnamefont {C.~W.}\ \bibnamefont
  {Zollitsch}}, \bibinfo {author} {\bibfnamefont {J.}~\bibnamefont {Lotze}},
  \bibinfo {author} {\bibfnamefont {F.}~\bibnamefont {Hocke}}, \bibinfo
  {author} {\bibfnamefont {M.}~\bibnamefont {Greifenstein}}, \bibinfo {author}
  {\bibfnamefont {A.}~\bibnamefont {Marx}}, \bibinfo {author} {\bibfnamefont
  {R.}~\bibnamefont {Gross}}, \ and\ \bibinfo {author} {\bibfnamefont
  {S.~T.~B.}\ \bibnamefont {Goennenwein}},\ }\href {\doibase
  10.1103/PhysRevLett.111.127003} {\bibfield  {journal} {\bibinfo  {journal}
  {Phys. Rev. Lett.}\ }\textbf {\bibinfo {volume} {111}},\ \bibinfo {pages}
  {127003} (\bibinfo {year} {2013})}\BibitemShut {NoStop}%
\bibitem [{\citenamefont {Tabuchi}\ \emph {et~al.}(2014)\citenamefont
  {Tabuchi}, \citenamefont {Ishino}, \citenamefont {Ishikawa}, \citenamefont
  {Yamazaki}, \citenamefont {Usami},\ and\ \citenamefont
  {Nakamura}}]{TabuchiPRL2014}%
  \BibitemOpen
  \bibfield  {author} {\bibinfo {author} {\bibfnamefont {Y.}~\bibnamefont
  {Tabuchi}}, \bibinfo {author} {\bibfnamefont {S.}~\bibnamefont {Ishino}},
  \bibinfo {author} {\bibfnamefont {T.}~\bibnamefont {Ishikawa}}, \bibinfo
  {author} {\bibfnamefont {R.}~\bibnamefont {Yamazaki}}, \bibinfo {author}
  {\bibfnamefont {K.}~\bibnamefont {Usami}}, \ and\ \bibinfo {author}
  {\bibfnamefont {Y.}~\bibnamefont {Nakamura}},\ }\href {\doibase
  10.1103/PhysRevLett.113.083603} {\bibfield  {journal} {\bibinfo  {journal}
  {Phys. Rev. Lett.}\ }\textbf {\bibinfo {volume} {113}},\ \bibinfo {pages}
  {083603} (\bibinfo {year} {2014})}\BibitemShut {NoStop}%
\bibitem [{\citenamefont {Zhang}\ \emph {et~al.}(2014)\citenamefont {Zhang},
  \citenamefont {Zou}, \citenamefont {Jiang},\ and\ \citenamefont
  {Tang}}]{ZhangPRL2014}%
  \BibitemOpen
  \bibfield  {author} {\bibinfo {author} {\bibfnamefont {X.}~\bibnamefont
  {Zhang}}, \bibinfo {author} {\bibfnamefont {C.-L.}\ \bibnamefont {Zou}},
  \bibinfo {author} {\bibfnamefont {L.}~\bibnamefont {Jiang}}, \ and\ \bibinfo
  {author} {\bibfnamefont {H.~X.}\ \bibnamefont {Tang}},\ }\href {\doibase
  10.1103/PhysRevLett.113.156401} {\bibfield  {journal} {\bibinfo  {journal}
  {Phys. Rev. Lett.}\ }\textbf {\bibinfo {volume} {113}},\ \bibinfo {pages}
  {156401} (\bibinfo {year} {2014})}\BibitemShut {NoStop}%
\bibitem [{\citenamefont {Bai}\ \emph {et~al.}(2015)\citenamefont {Bai},
  \citenamefont {Harder}, \citenamefont {Chen}, \citenamefont {Fan},
  \citenamefont {Xiao},\ and\ \citenamefont {Hu}}]{BaiPRL2015}%
  \BibitemOpen
  \bibfield  {author} {\bibinfo {author} {\bibfnamefont {L.}~\bibnamefont
  {Bai}}, \bibinfo {author} {\bibfnamefont {M.}~\bibnamefont {Harder}},
  \bibinfo {author} {\bibfnamefont {Y.~P.}\ \bibnamefont {Chen}}, \bibinfo
  {author} {\bibfnamefont {X.}~\bibnamefont {Fan}}, \bibinfo {author}
  {\bibfnamefont {J.~Q.}\ \bibnamefont {Xiao}}, \ and\ \bibinfo {author}
  {\bibfnamefont {C.-M.}\ \bibnamefont {Hu}},\ }\href {\doibase
  10.1103/PhysRevLett.114.227201} {\bibfield  {journal} {\bibinfo  {journal}
  {Phys. Rev. Lett.}\ }\textbf {\bibinfo {volume} {114}},\ \bibinfo {pages}
  {227201} (\bibinfo {year} {2015})}\BibitemShut {NoStop}%
\bibitem [{\citenamefont {Kuzmichev~et al.}(2020)}]{KuzmichevJETP2020}%
  \BibitemOpen
  \bibfield  {author} {\bibinfo {author} {\bibfnamefont {A.~N.}\ \bibnamefont
  {Kuzmichev~et al.}},\ }\href@noop {} {\bibfield  {journal} {\bibinfo
  {journal} {JETP Lett.}\ }\textbf {\bibinfo {volume} {112}},\ \bibinfo {pages}
  {710} (\bibinfo {year} {2020})}\BibitemShut {NoStop}%
\bibitem [{\citenamefont {Chumak}\ \emph {et~al.}(2015)\citenamefont {Chumak},
  \citenamefont {Vasyuchka}, \citenamefont {Serga},\ and\ \citenamefont
  {Hillebrands}}]{ChumakNPhys2015}%
  \BibitemOpen
  \bibfield  {author} {\bibinfo {author} {\bibfnamefont {A.~V.}\ \bibnamefont
  {Chumak}}, \bibinfo {author} {\bibfnamefont {V.~I.}\ \bibnamefont
  {Vasyuchka}}, \bibinfo {author} {\bibfnamefont {A.~A.}\ \bibnamefont
  {Serga}}, \ and\ \bibinfo {author} {\bibfnamefont {B.}~\bibnamefont
  {Hillebrands}},\ }\href@noop {} {\bibfield  {journal} {\bibinfo  {journal}
  {Nature Physics}\ }\textbf {\bibinfo {volume} {11}},\ \bibinfo {pages} {453}
  (\bibinfo {year} {2015})}\BibitemShut {NoStop}%
\bibitem [{\citenamefont {Imamo\u{g}lu}(2009)}]{ImamogluPRL2009}%
  \BibitemOpen
  \bibfield  {author} {\bibinfo {author} {\bibfnamefont {A.}~\bibnamefont
  {Imamo\u{g}lu}},\ }\href {\doibase 10.1103/PhysRevLett.102.083602} {\bibfield
   {journal} {\bibinfo  {journal} {Phys. Rev. Lett.}\ }\textbf {\bibinfo
  {volume} {102}},\ \bibinfo {pages} {083602} (\bibinfo {year}
  {2009})}\BibitemShut {NoStop}%
\bibitem [{\citenamefont {Wesenberg}\ \emph {et~al.}(2009)\citenamefont
  {Wesenberg}, \citenamefont {Ardavan}, \citenamefont {Briggs}, \citenamefont
  {Morton}, \citenamefont {Schoelkopf}, \citenamefont {Schuster},\ and\
  \citenamefont {M\o{}lmer}}]{WesenbergPRL2009}%
  \BibitemOpen
  \bibfield  {author} {\bibinfo {author} {\bibfnamefont {J.~H.}\ \bibnamefont
  {Wesenberg}}, \bibinfo {author} {\bibfnamefont {A.}~\bibnamefont {Ardavan}},
  \bibinfo {author} {\bibfnamefont {G.~A.~D.}\ \bibnamefont {Briggs}}, \bibinfo
  {author} {\bibfnamefont {J.~J.~L.}\ \bibnamefont {Morton}}, \bibinfo {author}
  {\bibfnamefont {R.~J.}\ \bibnamefont {Schoelkopf}}, \bibinfo {author}
  {\bibfnamefont {D.~I.}\ \bibnamefont {Schuster}}, \ and\ \bibinfo {author}
  {\bibfnamefont {K.}~\bibnamefont {M\o{}lmer}},\ }\href {\doibase
  10.1103/PhysRevLett.103.070502} {\bibfield  {journal} {\bibinfo  {journal}
  {Phys. Rev. Lett.}\ }\textbf {\bibinfo {volume} {103}},\ \bibinfo {pages}
  {070502} (\bibinfo {year} {2009})}\BibitemShut {NoStop}%
\bibitem [{\citenamefont {Tabuchi}\ \emph {et~al.}(2015)\citenamefont
  {Tabuchi}, \citenamefont {Ishino}, \citenamefont {Noguchi}, \citenamefont
  {Ishikawa}, \citenamefont {Yamazaki}, \citenamefont {Usami},\ and\
  \citenamefont {Nakamura}}]{TabuchiScience2015}%
  \BibitemOpen
  \bibfield  {author} {\bibinfo {author} {\bibfnamefont {Y.}~\bibnamefont
  {Tabuchi}}, \bibinfo {author} {\bibfnamefont {S.}~\bibnamefont {Ishino}},
  \bibinfo {author} {\bibfnamefont {A.}~\bibnamefont {Noguchi}}, \bibinfo
  {author} {\bibfnamefont {T.}~\bibnamefont {Ishikawa}}, \bibinfo {author}
  {\bibfnamefont {R.}~\bibnamefont {Yamazaki}}, \bibinfo {author}
  {\bibfnamefont {K.}~\bibnamefont {Usami}}, \ and\ \bibinfo {author}
  {\bibfnamefont {Y.}~\bibnamefont {Nakamura}},\ }\href@noop {} {\bibfield
  {journal} {\bibinfo  {journal} {Science}\ }\textbf {\bibinfo {volume}
  {349}},\ \bibinfo {pages} {405} (\bibinfo {year} {2015})}\BibitemShut
  {NoStop}%
\bibitem [{\citenamefont {Lachance-Quirion}\ \emph {et~al.}(2020)\citenamefont
  {Lachance-Quirion}, \citenamefont {Wolski}, \citenamefont {Tabuchi},
  \citenamefont {Kono}, \citenamefont {Usami},\ and\ \citenamefont
  {Nakamura}}]{LachanceQuirionScience2020}%
  \BibitemOpen
  \bibfield  {author} {\bibinfo {author} {\bibfnamefont {D.}~\bibnamefont
  {Lachance-Quirion}}, \bibinfo {author} {\bibfnamefont {S.~P.}\ \bibnamefont
  {Wolski}}, \bibinfo {author} {\bibfnamefont {Y.}~\bibnamefont {Tabuchi}},
  \bibinfo {author} {\bibfnamefont {S.}~\bibnamefont {Kono}}, \bibinfo {author}
  {\bibfnamefont {K.}~\bibnamefont {Usami}}, \ and\ \bibinfo {author}
  {\bibfnamefont {Y.}~\bibnamefont {Nakamura}},\ }\href@noop {} {\bibfield
  {journal} {\bibinfo  {journal} {Science}\ }\textbf {\bibinfo {volume}
  {367}},\ \bibinfo {pages} {425} (\bibinfo {year} {2020})}\BibitemShut
  {NoStop}%
\bibitem [{\citenamefont {Trifunovic}\ \emph {et~al.}(2013)\citenamefont
  {Trifunovic}, \citenamefont {Pedrocchi},\ and\ \citenamefont
  {Loss}}]{TrifunovicPRX2013}%
  \BibitemOpen
  \bibfield  {author} {\bibinfo {author} {\bibfnamefont {L.}~\bibnamefont
  {Trifunovic}}, \bibinfo {author} {\bibfnamefont {F.~L.}\ \bibnamefont
  {Pedrocchi}}, \ and\ \bibinfo {author} {\bibfnamefont {D.}~\bibnamefont
  {Loss}},\ }\href@noop {} {\bibfield  {journal} {\bibinfo  {journal} {Phys.
  Rev. X}\ }\textbf {\bibinfo {volume} {3}},\ \bibinfo {pages} {041023}
  (\bibinfo {year} {2013})}\BibitemShut {NoStop}%
\bibitem [{\citenamefont {Andrich}\ \emph {et~al.}(2017)\citenamefont
  {Andrich}, \citenamefont {de~las Casas}, \citenamefont {Liu}, \citenamefont
  {Bretscher}, \citenamefont {Berman}, \citenamefont {Heremans}, \citenamefont
  {Nealey},\ and\ \citenamefont {Awschalom}}]{AndrichnpjQuantumInf2017}%
  \BibitemOpen
  \bibfield  {author} {\bibinfo {author} {\bibfnamefont {P.}~\bibnamefont
  {Andrich}}, \bibinfo {author} {\bibfnamefont {C.~F.}\ \bibnamefont {de~las
  Casas}}, \bibinfo {author} {\bibfnamefont {X.}~\bibnamefont {Liu}}, \bibinfo
  {author} {\bibfnamefont {H.~L.}\ \bibnamefont {Bretscher}}, \bibinfo {author}
  {\bibfnamefont {J.~R.}\ \bibnamefont {Berman}}, \bibinfo {author}
  {\bibfnamefont {F.~J.}\ \bibnamefont {Heremans}}, \bibinfo {author}
  {\bibfnamefont {P.~F.}\ \bibnamefont {Nealey}}, \ and\ \bibinfo {author}
  {\bibfnamefont {D.~D.}\ \bibnamefont {Awschalom}},\ }\href@noop {} {\bibfield
   {journal} {\bibinfo  {journal} {npj Quantum Inf.}\ }\textbf {\bibinfo
  {volume} {3}},\ \bibinfo {pages} {28} (\bibinfo {year} {2017})}\BibitemShut
  {NoStop}%
\bibitem [{\citenamefont {Rusconi}\ \emph {et~al.}(2019)\citenamefont
  {Rusconi}, \citenamefont {Schuetz}, \citenamefont {Gieseler}, \citenamefont
  {Lukin},\ and\ \citenamefont {Romero-Isart}}]{RusconiPRA2019}%
  \BibitemOpen
  \bibfield  {author} {\bibinfo {author} {\bibfnamefont {C.~C.}\ \bibnamefont
  {Rusconi}}, \bibinfo {author} {\bibfnamefont {M.~J.~A.}\ \bibnamefont
  {Schuetz}}, \bibinfo {author} {\bibfnamefont {J.}~\bibnamefont {Gieseler}},
  \bibinfo {author} {\bibfnamefont {M.~D.}\ \bibnamefont {Lukin}}, \ and\
  \bibinfo {author} {\bibfnamefont {O.}~\bibnamefont {Romero-Isart}},\ }\href
  {\doibase 10.1103/PhysRevA.100.022343} {\bibfield  {journal} {\bibinfo
  {journal} {Phys. Rev. A}\ }\textbf {\bibinfo {volume} {100}},\ \bibinfo
  {pages} {022343} (\bibinfo {year} {2019})}\BibitemShut {NoStop}%
\bibitem [{\citenamefont {Elyasi}\ \emph {et~al.}(2020)\citenamefont {Elyasi},
  \citenamefont {Blanter},\ and\ \citenamefont {Bauer}}]{ElyasiPRB2020}%
  \BibitemOpen
  \bibfield  {author} {\bibinfo {author} {\bibfnamefont {M.}~\bibnamefont
  {Elyasi}}, \bibinfo {author} {\bibfnamefont {Y.~M.}\ \bibnamefont {Blanter}},
  \ and\ \bibinfo {author} {\bibfnamefont {G.~E.~W.}\ \bibnamefont {Bauer}},\
  }\href {\doibase 10.1103/PhysRevB.101.054402} {\bibfield  {journal} {\bibinfo
   {journal} {Phys. Rev. B}\ }\textbf {\bibinfo {volume} {101}},\ \bibinfo
  {pages} {054402} (\bibinfo {year} {2020})}\BibitemShut {NoStop}%
\bibitem [{\citenamefont {Neuman}\ \emph {et~al.}(2020)\citenamefont {Neuman},
  \citenamefont {Wang},\ and\ \citenamefont {Narang}}]{NeumanPRL2020}%
  \BibitemOpen
  \bibfield  {author} {\bibinfo {author} {\bibfnamefont {T.~c.~v.}\
  \bibnamefont {Neuman}}, \bibinfo {author} {\bibfnamefont {D.~S.}\
  \bibnamefont {Wang}}, \ and\ \bibinfo {author} {\bibfnamefont
  {P.}~\bibnamefont {Narang}},\ }\href {\doibase
  10.1103/PhysRevLett.125.247702} {\bibfield  {journal} {\bibinfo  {journal}
  {Phys. Rev. Lett.}\ }\textbf {\bibinfo {volume} {125}},\ \bibinfo {pages}
  {247702} (\bibinfo {year} {2020})}\BibitemShut {NoStop}%
\bibitem [{\citenamefont {Morris}\ \emph {et~al.}(2017)\citenamefont {Morris},
  \citenamefont {van Loo}, \citenamefont {Kosen},\ and\ \citenamefont
  {Karenowska}}]{MorrisSREP2017}%
  \BibitemOpen
  \bibfield  {author} {\bibinfo {author} {\bibfnamefont {R.~G.~E.}\
  \bibnamefont {Morris}}, \bibinfo {author} {\bibfnamefont {A.~F.}\
  \bibnamefont {van Loo}}, \bibinfo {author} {\bibfnamefont {S.}~\bibnamefont
  {Kosen}}, \ and\ \bibinfo {author} {\bibfnamefont {A.~D.}\ \bibnamefont
  {Karenowska}},\ }\href@noop {} {\bibfield  {journal} {\bibinfo  {journal}
  {Sci. Rep.}\ }\textbf {\bibinfo {volume} {7}},\ \bibinfo {pages} {11511}
  (\bibinfo {year} {2017})}\BibitemShut {NoStop}%
\bibitem [{\citenamefont {Mandal}\ \emph {et~al.}(2020)\citenamefont {Mandal},
  \citenamefont {Kapoor}, \citenamefont {Ghosh}, \citenamefont {Jesudasan},
  \citenamefont {Manni}, \citenamefont {Thamizhavel}, \citenamefont
  {Raychaudhuri}, \citenamefont {Singh},\ and\ \citenamefont
  {Deshmukh}}]{MandalAPL2020}%
  \BibitemOpen
  \bibfield  {author} {\bibinfo {author} {\bibfnamefont {S.}~\bibnamefont
  {Mandal}}, \bibinfo {author} {\bibfnamefont {L.~N.}\ \bibnamefont {Kapoor}},
  \bibinfo {author} {\bibfnamefont {S.}~\bibnamefont {Ghosh}}, \bibinfo
  {author} {\bibfnamefont {J.}~\bibnamefont {Jesudasan}}, \bibinfo {author}
  {\bibfnamefont {S.}~\bibnamefont {Manni}}, \bibinfo {author} {\bibfnamefont
  {A.}~\bibnamefont {Thamizhavel}}, \bibinfo {author} {\bibfnamefont
  {P.}~\bibnamefont {Raychaudhuri}}, \bibinfo {author} {\bibfnamefont
  {V.}~\bibnamefont {Singh}}, \ and\ \bibinfo {author} {\bibfnamefont {M.~M.}\
  \bibnamefont {Deshmukh}},\ }\href@noop {} {\bibfield  {journal} {\bibinfo
  {journal} {Appl. Phys. Lett.}\ }\textbf {\bibinfo {volume} {117}},\ \bibinfo
  {pages} {263101} (\bibinfo {year} {2020})}\BibitemShut {NoStop}%
\bibitem [{\citenamefont {Haygood}\ \emph {et~al.}(2021)\citenamefont
  {Haygood}, \citenamefont {Pufall}, \citenamefont {Edwards}, \citenamefont
  {Shaw},\ and\ \citenamefont {Rippard}}]{HaygoodPRApplied2021}%
  \BibitemOpen
  \bibfield  {author} {\bibinfo {author} {\bibfnamefont {I.}~\bibnamefont
  {Haygood}}, \bibinfo {author} {\bibfnamefont {M.}~\bibnamefont {Pufall}},
  \bibinfo {author} {\bibfnamefont {E.}~\bibnamefont {Edwards}}, \bibinfo
  {author} {\bibfnamefont {J.~M.}\ \bibnamefont {Shaw}}, \ and\ \bibinfo
  {author} {\bibfnamefont {W.}~\bibnamefont {Rippard}},\ }\href {\doibase
  10.1103/PhysRevApplied.15.054021} {\bibfield  {journal} {\bibinfo  {journal}
  {Phys. Rev. Applied}\ }\textbf {\bibinfo {volume} {15}},\ \bibinfo {pages}
  {054021} (\bibinfo {year} {2021})}\BibitemShut {NoStop}%
\bibitem [{\citenamefont {Baity}\ \emph {et~al.}(2021)\citenamefont {Baity},
  \citenamefont {Bozhko}, \citenamefont {Macêdo}, \citenamefont {Smith},
  \citenamefont {Holland}, \citenamefont {Danilin}, \citenamefont {Seferai},
  \citenamefont {Barbosa}, \citenamefont {Peroor}, \citenamefont {Goldman},
  \citenamefont {Nasti}, \citenamefont {Paul}, \citenamefont {Hadfield},
  \citenamefont {McVitie},\ and\ \citenamefont {Weides}}]{BaityAPL2021}%
  \BibitemOpen
  \bibfield  {author} {\bibinfo {author} {\bibfnamefont {P.~G.}\ \bibnamefont
  {Baity}}, \bibinfo {author} {\bibfnamefont {D.~A.}\ \bibnamefont {Bozhko}},
  \bibinfo {author} {\bibfnamefont {R.}~\bibnamefont {Macêdo}}, \bibinfo
  {author} {\bibfnamefont {W.}~\bibnamefont {Smith}}, \bibinfo {author}
  {\bibfnamefont {R.~C.}\ \bibnamefont {Holland}}, \bibinfo {author}
  {\bibfnamefont {S.}~\bibnamefont {Danilin}}, \bibinfo {author} {\bibfnamefont
  {V.}~\bibnamefont {Seferai}}, \bibinfo {author} {\bibfnamefont
  {J.}~\bibnamefont {Barbosa}}, \bibinfo {author} {\bibfnamefont {R.~R.}\
  \bibnamefont {Peroor}}, \bibinfo {author} {\bibfnamefont {S.}~\bibnamefont
  {Goldman}}, \bibinfo {author} {\bibfnamefont {U.}~\bibnamefont {Nasti}},
  \bibinfo {author} {\bibfnamefont {J.}~\bibnamefont {Paul}}, \bibinfo {author}
  {\bibfnamefont {R.~H.}\ \bibnamefont {Hadfield}}, \bibinfo {author}
  {\bibfnamefont {S.}~\bibnamefont {McVitie}}, \ and\ \bibinfo {author}
  {\bibfnamefont {M.}~\bibnamefont {Weides}},\ }\href@noop {} {\bibfield
  {journal} {\bibinfo  {journal} {Appl. Phys. Lett.}\ }\textbf {\bibinfo
  {volume} {119}},\ \bibinfo {pages} {033502} (\bibinfo {year}
  {2021})}\BibitemShut {NoStop}%
\bibitem [{\citenamefont {Heyroth}\ \emph {et~al.}(2019)\citenamefont
  {Heyroth}, \citenamefont {Hauser}, \citenamefont {Trempler}, \citenamefont
  {Geyer}, \citenamefont {Syrowatka}, \citenamefont {Dreyer}, \citenamefont
  {Ebbinghaus}, \citenamefont {Woltersdorf},\ and\ \citenamefont
  {Schmidt}}]{HeyrothPRApplied2019}%
  \BibitemOpen
  \bibfield  {author} {\bibinfo {author} {\bibfnamefont {F.}~\bibnamefont
  {Heyroth}}, \bibinfo {author} {\bibfnamefont {C.}~\bibnamefont {Hauser}},
  \bibinfo {author} {\bibfnamefont {P.}~\bibnamefont {Trempler}}, \bibinfo
  {author} {\bibfnamefont {P.}~\bibnamefont {Geyer}}, \bibinfo {author}
  {\bibfnamefont {F.}~\bibnamefont {Syrowatka}}, \bibinfo {author}
  {\bibfnamefont {R.}~\bibnamefont {Dreyer}}, \bibinfo {author} {\bibfnamefont
  {S.}~\bibnamefont {Ebbinghaus}}, \bibinfo {author} {\bibfnamefont
  {G.}~\bibnamefont {Woltersdorf}}, \ and\ \bibinfo {author} {\bibfnamefont
  {G.}~\bibnamefont {Schmidt}},\ }\href {\doibase
  10.1103/PhysRevApplied.12.054031} {\bibfield  {journal} {\bibinfo  {journal}
  {Phys. Rev. Applied}\ }\textbf {\bibinfo {volume} {12}},\ \bibinfo {pages}
  {054031} (\bibinfo {year} {2019})}\BibitemShut {NoStop}%
\bibitem [{\citenamefont {Trempler}\ \emph {et~al.}(2020)\citenamefont
  {Trempler}, \citenamefont {Dreyer}, \citenamefont {Geyer}, \citenamefont
  {Hauser}, \citenamefont {Woltersdorf},\ and\ \citenamefont
  {Schmidt}}]{TremplerAPL2020}%
  \BibitemOpen
  \bibfield  {author} {\bibinfo {author} {\bibfnamefont {P.}~\bibnamefont
  {Trempler}}, \bibinfo {author} {\bibfnamefont {R.}~\bibnamefont {Dreyer}},
  \bibinfo {author} {\bibfnamefont {P.}~\bibnamefont {Geyer}}, \bibinfo
  {author} {\bibfnamefont {C.}~\bibnamefont {Hauser}}, \bibinfo {author}
  {\bibfnamefont {G.}~\bibnamefont {Woltersdorf}}, \ and\ \bibinfo {author}
  {\bibfnamefont {G.}~\bibnamefont {Schmidt}},\ }\href@noop {} {\bibfield
  {journal} {\bibinfo  {journal} {Appl. Phys. Lett.}\ }\textbf {\bibinfo
  {volume} {117}},\ \bibinfo {pages} {232401} (\bibinfo {year}
  {2020})}\BibitemShut {NoStop}%
\bibitem [{\citenamefont {Kosen}\ \emph {et~al.}(2019)\citenamefont {Kosen},
  \citenamefont {van Loo}, \citenamefont {Bozhko}, \citenamefont {Mihalceanu},\
  and\ \citenamefont {Karenowska}}]{KosenAPLMaterials2019}%
  \BibitemOpen
  \bibfield  {author} {\bibinfo {author} {\bibfnamefont {S.}~\bibnamefont
  {Kosen}}, \bibinfo {author} {\bibfnamefont {A.~F.}\ \bibnamefont {van Loo}},
  \bibinfo {author} {\bibfnamefont {D.~A.}\ \bibnamefont {Bozhko}}, \bibinfo
  {author} {\bibfnamefont {L.}~\bibnamefont {Mihalceanu}}, \ and\ \bibinfo
  {author} {\bibfnamefont {A.~D.}\ \bibnamefont {Karenowska}},\ }\href@noop {}
  {\bibfield  {journal} {\bibinfo  {journal} {APL Materials}\ }\textbf
  {\bibinfo {volume} {7}},\ \bibinfo {pages} {101120} (\bibinfo {year}
  {2019})}\BibitemShut {NoStop}%
\bibitem [{\citenamefont {Harder}\ \emph {et~al.}(2018)\citenamefont {Harder},
  \citenamefont {Yang}, \citenamefont {Yao}, \citenamefont {Yu}, \citenamefont
  {Rao}, \citenamefont {Gui}, \citenamefont {Stamps},\ and\ \citenamefont
  {Hu}}]{HarderPRL2018}%
  \BibitemOpen
  \bibfield  {author} {\bibinfo {author} {\bibfnamefont {M.}~\bibnamefont
  {Harder}}, \bibinfo {author} {\bibfnamefont {Y.}~\bibnamefont {Yang}},
  \bibinfo {author} {\bibfnamefont {B.~M.}\ \bibnamefont {Yao}}, \bibinfo
  {author} {\bibfnamefont {C.~H.}\ \bibnamefont {Yu}}, \bibinfo {author}
  {\bibfnamefont {J.~W.}\ \bibnamefont {Rao}}, \bibinfo {author} {\bibfnamefont
  {Y.~S.}\ \bibnamefont {Gui}}, \bibinfo {author} {\bibfnamefont {R.~L.}\
  \bibnamefont {Stamps}}, \ and\ \bibinfo {author} {\bibfnamefont {C.-M.}\
  \bibnamefont {Hu}},\ }\href {\doibase 10.1103/PhysRevLett.121.137203}
  {\bibfield  {journal} {\bibinfo  {journal} {Phys. Rev. Lett.}\ }\textbf
  {\bibinfo {volume} {121}},\ \bibinfo {pages} {137203} (\bibinfo {year}
  {2018})}\BibitemShut {NoStop}%
\bibitem [{\citenamefont {Bhoi}\ \emph {et~al.}(2019)\citenamefont {Bhoi},
  \citenamefont {Kim}, \citenamefont {Jang}, \citenamefont {Kim}, \citenamefont
  {Yang}, \citenamefont {Cho},\ and\ \citenamefont {Kim}}]{BhoiPRB2019}%
  \BibitemOpen
  \bibfield  {author} {\bibinfo {author} {\bibfnamefont {B.}~\bibnamefont
  {Bhoi}}, \bibinfo {author} {\bibfnamefont {B.}~\bibnamefont {Kim}}, \bibinfo
  {author} {\bibfnamefont {S.-H.}\ \bibnamefont {Jang}}, \bibinfo {author}
  {\bibfnamefont {J.}~\bibnamefont {Kim}}, \bibinfo {author} {\bibfnamefont
  {J.}~\bibnamefont {Yang}}, \bibinfo {author} {\bibfnamefont {Y.-J.}\
  \bibnamefont {Cho}}, \ and\ \bibinfo {author} {\bibfnamefont {S.-K.}\
  \bibnamefont {Kim}},\ }\href {\doibase 10.1103/PhysRevB.99.134426} {\bibfield
   {journal} {\bibinfo  {journal} {Phys. Rev. B}\ }\textbf {\bibinfo {volume}
  {99}},\ \bibinfo {pages} {134426} (\bibinfo {year} {2019})}\BibitemShut
  {NoStop}%
\bibitem [{\citenamefont {Boventer}\ \emph {et~al.}(2020)\citenamefont
  {Boventer}, \citenamefont {D\"orflinger}, \citenamefont {Wolz}, \citenamefont
  {Mac\^edo}, \citenamefont {Lebrun}, \citenamefont {Kl\"aui},\ and\
  \citenamefont {Weides}}]{BoventerPRR2020}%
  \BibitemOpen
  \bibfield  {author} {\bibinfo {author} {\bibfnamefont {I.}~\bibnamefont
  {Boventer}}, \bibinfo {author} {\bibfnamefont {C.}~\bibnamefont
  {D\"orflinger}}, \bibinfo {author} {\bibfnamefont {T.}~\bibnamefont {Wolz}},
  \bibinfo {author} {\bibfnamefont {R.}~\bibnamefont {Mac\^edo}}, \bibinfo
  {author} {\bibfnamefont {R.}~\bibnamefont {Lebrun}}, \bibinfo {author}
  {\bibfnamefont {M.}~\bibnamefont {Kl\"aui}}, \ and\ \bibinfo {author}
  {\bibfnamefont {M.}~\bibnamefont {Weides}},\ }\href {\doibase
  10.1103/PhysRevResearch.2.013154} {\bibfield  {journal} {\bibinfo  {journal}
  {Phys. Rev. Research}\ }\textbf {\bibinfo {volume} {2}},\ \bibinfo {pages}
  {013154} (\bibinfo {year} {2020})}\BibitemShut {NoStop}%
\bibitem [{\citenamefont {Tserkovnyak}(2020)}]{TserkovnyakPRR2020}%
  \BibitemOpen
  \bibfield  {author} {\bibinfo {author} {\bibfnamefont {Y.}~\bibnamefont
  {Tserkovnyak}},\ }\href {\doibase 10.1103/PhysRevResearch.2.013031}
  {\bibfield  {journal} {\bibinfo  {journal} {Phys. Rev. Research}\ }\textbf
  {\bibinfo {volume} {2}},\ \bibinfo {pages} {013031} (\bibinfo {year}
  {2020})}\BibitemShut {NoStop}%
\bibitem [{sup()}]{supplement}%
  \BibitemOpen
  \href@noop {} {\ }\bibinfo {note} {See the Supplemental Information for
  details.}\BibitemShut {Stop}%
\bibitem [{\citenamefont {Lambert}\ \emph {et~al.}(2016)\citenamefont
  {Lambert}, \citenamefont {Haigh}, \citenamefont {Langenfeld}, \citenamefont
  {Doherty},\ and\ \citenamefont {Ferguson}}]{LambertPRA2016}%
  \BibitemOpen
  \bibfield  {author} {\bibinfo {author} {\bibfnamefont {N.~J.}\ \bibnamefont
  {Lambert}}, \bibinfo {author} {\bibfnamefont {J.~A.}\ \bibnamefont {Haigh}},
  \bibinfo {author} {\bibfnamefont {S.}~\bibnamefont {Langenfeld}}, \bibinfo
  {author} {\bibfnamefont {A.~C.}\ \bibnamefont {Doherty}}, \ and\ \bibinfo
  {author} {\bibfnamefont {A.~J.}\ \bibnamefont {Ferguson}},\ }\href {\doibase
  10.1103/PhysRevA.93.021803} {\bibfield  {journal} {\bibinfo  {journal} {Phys.
  Rev. A}\ }\textbf {\bibinfo {volume} {93}},\ \bibinfo {pages} {021803}
  (\bibinfo {year} {2016})}\BibitemShut {NoStop}%
\bibitem [{\citenamefont {Zare~Rameshti}\ and\ \citenamefont
  {Bauer}(2018)}]{RameshtiPRB2018}%
  \BibitemOpen
  \bibfield  {author} {\bibinfo {author} {\bibfnamefont {B.}~\bibnamefont
  {Zare~Rameshti}}\ and\ \bibinfo {author} {\bibfnamefont {G.~E.~W.}\
  \bibnamefont {Bauer}},\ }\href {\doibase 10.1103/PhysRevB.97.014419}
  {\bibfield  {journal} {\bibinfo  {journal} {Phys. Rev. B}\ }\textbf {\bibinfo
  {volume} {97}},\ \bibinfo {pages} {014419} (\bibinfo {year}
  {2018})}\BibitemShut {NoStop}%
\bibitem [{\citenamefont {Grigoryan}\ and\ \citenamefont
  {Xia}(2019)}]{GrigoryanPRB2019}%
  \BibitemOpen
  \bibfield  {author} {\bibinfo {author} {\bibfnamefont {V.~L.}\ \bibnamefont
  {Grigoryan}}\ and\ \bibinfo {author} {\bibfnamefont {K.}~\bibnamefont
  {Xia}},\ }\href {\doibase 10.1103/PhysRevB.100.014415} {\bibfield  {journal}
  {\bibinfo  {journal} {Phys. Rev. B}\ }\textbf {\bibinfo {volume} {100}},\
  \bibinfo {pages} {014415} (\bibinfo {year} {2019})}\BibitemShut {NoStop}%
\bibitem [{\citenamefont {Xu}\ \emph {et~al.}(2019)\citenamefont {Xu},
  \citenamefont {Rao}, \citenamefont {Gui}, \citenamefont {Jin},\ and\
  \citenamefont {Hu}}]{XuPRB2019}%
  \BibitemOpen
  \bibfield  {author} {\bibinfo {author} {\bibfnamefont {P.-C.}\ \bibnamefont
  {Xu}}, \bibinfo {author} {\bibfnamefont {J.~W.}\ \bibnamefont {Rao}},
  \bibinfo {author} {\bibfnamefont {Y.~S.}\ \bibnamefont {Gui}}, \bibinfo
  {author} {\bibfnamefont {X.}~\bibnamefont {Jin}}, \ and\ \bibinfo {author}
  {\bibfnamefont {C.-M.}\ \bibnamefont {Hu}},\ }\href {\doibase
  10.1103/PhysRevB.100.094415} {\bibfield  {journal} {\bibinfo  {journal}
  {Phys. Rev. B}\ }\textbf {\bibinfo {volume} {100}},\ \bibinfo {pages}
  {094415} (\bibinfo {year} {2019})}\BibitemShut {NoStop}%
\bibitem [{\citenamefont {Zhang}\ \emph {et~al.}(2015)\citenamefont {Zhang},
  \citenamefont {Zou}, \citenamefont {Zhu}, \citenamefont {Marquardt},
  \citenamefont {Jiang},\ and\ \citenamefont {Tang}}]{ZhangNComm2015}%
  \BibitemOpen
  \bibfield  {author} {\bibinfo {author} {\bibfnamefont {X.}~\bibnamefont
  {Zhang}}, \bibinfo {author} {\bibfnamefont {C.-L.}\ \bibnamefont {Zou}},
  \bibinfo {author} {\bibfnamefont {N.}~\bibnamefont {Zhu}}, \bibinfo {author}
  {\bibfnamefont {F.}~\bibnamefont {Marquardt}}, \bibinfo {author}
  {\bibfnamefont {L.}~\bibnamefont {Jiang}}, \ and\ \bibinfo {author}
  {\bibfnamefont {H.~X.}\ \bibnamefont {Tang}},\ }\href@noop {} {\bibfield
  {journal} {\bibinfo  {journal} {Nature Communi.}\ }\textbf {\bibinfo {volume}
  {6}},\ \bibinfo {pages} {8914} (\bibinfo {year} {2015})}\BibitemShut
  {NoStop}%
\bibitem [{\citenamefont {Yao}\ \emph {et~al.}(2019)\citenamefont {Yao},
  \citenamefont {Yu}, \citenamefont {Zhang}, \citenamefont {Lu}, \citenamefont
  {Gui}, \citenamefont {Hu},\ and\ \citenamefont {Blanter}}]{YaoPRB2019}%
  \BibitemOpen
  \bibfield  {author} {\bibinfo {author} {\bibfnamefont {B.}~\bibnamefont
  {Yao}}, \bibinfo {author} {\bibfnamefont {T.}~\bibnamefont {Yu}}, \bibinfo
  {author} {\bibfnamefont {X.}~\bibnamefont {Zhang}}, \bibinfo {author}
  {\bibfnamefont {W.}~\bibnamefont {Lu}}, \bibinfo {author} {\bibfnamefont
  {Y.}~\bibnamefont {Gui}}, \bibinfo {author} {\bibfnamefont {C.-M.}\
  \bibnamefont {Hu}}, \ and\ \bibinfo {author} {\bibfnamefont {Y.~M.}\
  \bibnamefont {Blanter}},\ }\href {\doibase 10.1103/PhysRevB.100.214426}
  {\bibfield  {journal} {\bibinfo  {journal} {Phys. Rev. B}\ }\textbf {\bibinfo
  {volume} {100}},\ \bibinfo {pages} {214426} (\bibinfo {year}
  {2019})}\BibitemShut {NoStop}%
\bibitem [{\citenamefont {Yu}\ \emph {et~al.}(2020)\citenamefont {Yu},
  \citenamefont {Zhang}, \citenamefont {Sharma}, \citenamefont {Blanter},\ and\
  \citenamefont {Bauer}}]{YuPRB2020}%
  \BibitemOpen
  \bibfield  {author} {\bibinfo {author} {\bibfnamefont {T.}~\bibnamefont
  {Yu}}, \bibinfo {author} {\bibfnamefont {X.}~\bibnamefont {Zhang}}, \bibinfo
  {author} {\bibfnamefont {S.}~\bibnamefont {Sharma}}, \bibinfo {author}
  {\bibfnamefont {Y.~M.}\ \bibnamefont {Blanter}}, \ and\ \bibinfo {author}
  {\bibfnamefont {G.~E.~W.}\ \bibnamefont {Bauer}},\ }\href {\doibase
  10.1103/PhysRevB.101.094414} {\bibfield  {journal} {\bibinfo  {journal}
  {Phys. Rev. B}\ }\textbf {\bibinfo {volume} {101}},\ \bibinfo {pages}
  {094414} (\bibinfo {year} {2020})}\BibitemShut {NoStop}%
\bibitem [{\citenamefont {Trickle}\ \emph {et~al.}(2020)\citenamefont
  {Trickle}, \citenamefont {Zhang},\ and\ \citenamefont
  {Zurek}}]{TricklePRL2020}%
  \BibitemOpen
  \bibfield  {author} {\bibinfo {author} {\bibfnamefont {T.}~\bibnamefont
  {Trickle}}, \bibinfo {author} {\bibfnamefont {Z.}~\bibnamefont {Zhang}}, \
  and\ \bibinfo {author} {\bibfnamefont {K.~M.}\ \bibnamefont {Zurek}},\ }\href
  {\doibase 10.1103/PhysRevLett.124.201801} {\bibfield  {journal} {\bibinfo
  {journal} {Phys. Rev. Lett.}\ }\textbf {\bibinfo {volume} {124}},\ \bibinfo
  {pages} {201801} (\bibinfo {year} {2020})}\BibitemShut {NoStop}%
\bibitem [{\citenamefont {Heinz}\ \emph {et~al.}(2020)\citenamefont {Heinz},
  \citenamefont {Br\"{a}cher}, \citenamefont {Schneider}, \citenamefont {Wang},
  \citenamefont {L\"{a}gel}, \citenamefont {Friedel}, \citenamefont
  {Breitbach}, \citenamefont {Steinert}, \citenamefont {Meyer}, \citenamefont
  {Kewenig}, \citenamefont {Dubs}, \citenamefont {Pirro},\ and\ \citenamefont
  {Chumak}}]{HeinzNanoLett2020}%
  \BibitemOpen
  \bibfield  {author} {\bibinfo {author} {\bibfnamefont {B.}~\bibnamefont
  {Heinz}}, \bibinfo {author} {\bibfnamefont {T.}~\bibnamefont {Br\"{a}cher}},
  \bibinfo {author} {\bibfnamefont {M.}~\bibnamefont {Schneider}}, \bibinfo
  {author} {\bibfnamefont {Q.}~\bibnamefont {Wang}}, \bibinfo {author}
  {\bibfnamefont {B.}~\bibnamefont {L\"{a}gel}}, \bibinfo {author}
  {\bibfnamefont {A.~M.}\ \bibnamefont {Friedel}}, \bibinfo {author}
  {\bibfnamefont {D.}~\bibnamefont {Breitbach}}, \bibinfo {author}
  {\bibfnamefont {S.}~\bibnamefont {Steinert}}, \bibinfo {author}
  {\bibfnamefont {T.}~\bibnamefont {Meyer}}, \bibinfo {author} {\bibfnamefont
  {M.}~\bibnamefont {Kewenig}}, \bibinfo {author} {\bibfnamefont
  {C.}~\bibnamefont {Dubs}}, \bibinfo {author} {\bibfnamefont {P.}~\bibnamefont
  {Pirro}}, \ and\ \bibinfo {author} {\bibfnamefont {A.~V.}\ \bibnamefont
  {Chumak}},\ }\href@noop {} {\bibfield  {journal} {\bibinfo  {journal} {Nano
  Lett.}\ }\textbf {\bibinfo {volume} {20}},\ \bibinfo {pages} {4220} (\bibinfo
  {year} {2020})}\BibitemShut {NoStop}%
\bibitem [{\citenamefont {Fan}\ \emph {et~al.}(2020)\citenamefont {Fan},
  \citenamefont {Quarterman}, \citenamefont {Finley}, \citenamefont {Han},
  \citenamefont {Zhang}, \citenamefont {Hou}, \citenamefont {Stiles},
  \citenamefont {Grutter},\ and\ \citenamefont {Liu}}]{FanPRApplied2020}%
  \BibitemOpen
  \bibfield  {author} {\bibinfo {author} {\bibfnamefont {Y.}~\bibnamefont
  {Fan}}, \bibinfo {author} {\bibfnamefont {P.}~\bibnamefont {Quarterman}},
  \bibinfo {author} {\bibfnamefont {J.}~\bibnamefont {Finley}}, \bibinfo
  {author} {\bibfnamefont {J.}~\bibnamefont {Han}}, \bibinfo {author}
  {\bibfnamefont {P.}~\bibnamefont {Zhang}}, \bibinfo {author} {\bibfnamefont
  {J.~T.}\ \bibnamefont {Hou}}, \bibinfo {author} {\bibfnamefont {M.~D.}\
  \bibnamefont {Stiles}}, \bibinfo {author} {\bibfnamefont {A.~J.}\
  \bibnamefont {Grutter}}, \ and\ \bibinfo {author} {\bibfnamefont
  {L.}~\bibnamefont {Liu}},\ }\href {\doibase 10.1103/PhysRevApplied.13.061002}
  {\bibfield  {journal} {\bibinfo  {journal} {Phys. Rev. Applied}\ }\textbf
  {\bibinfo {volume} {13}},\ \bibinfo {pages} {061002} (\bibinfo {year}
  {2020})}\BibitemShut {NoStop}%
\end{thebibliography}

%

\end{document}